\documentclass[prd,preprintnumbers,twocolumn,eqsecnum,floatfix,a4paper,nofootinbib,superscriptaddress]{revtex4-1}

\usepackage{color}
\usepackage{amsmath,amssymb,graphicx,mathtools}
\usepackage{bm,lipsum}
\usepackage{times}
\usepackage{microtype}
\usepackage[utf8]{inputenc}
\usepackage{booktabs}
\usepackage[caption=false]{subfig}
\usepackage[normalem]{ulem}
\usepackage[varg]{txfonts}
\usepackage{bm,graphics,graphicx,epsfig,xcolor,ulem}
\usepackage[colorlinks, pdfborder={0 0 0}]{hyperref}
\usepackage{multirow}
\definecolor{LinkColor}{rgb}{0.75, 0, 0}
\definecolor{CiteColor}{rgb}{0, 0.5, 0.5}
\definecolor{UrlColor}{rgb}{0, 0, 0.75}
\hypersetup{linkcolor=LinkColor}
\hypersetup{citecolor=CiteColor}
\hypersetup{urlcolor=UrlColor}
\allowdisplaybreaks
\begin{document}
\title{Importance of mirror modes in binary black hole ringdown waveform}


\newcommand{\penncosmos}{\affiliation{Institute for Gravitation and the Cosmos, Department of Physics, Pennsylvania State University, University Park, PA, 16802, USA}}
\newcommand{\pennastro}{\affiliation{Department of Astronomy \& Astrophysics, Pennsylvania State University, University Park, PA, 16802, USA}}
\newcommand{\cardiff}{\affiliation{School of Physics and Astronomy, Cardiff University, Cardiff, UK, CF24 3AA}
}


\author{Arnab Dhani}
\email{aud371@psu.edu}
\penncosmos 

\begin{abstract}
    The post-merger signal in binary black hole merger is described by linear, black-hole perturbation theory.
    Historically, this has been modeled using the dominant positive-frequency (corotating), fundamental mode. 
    Recently, there has been a renewed effort in modeling the post-merger waveform using higher, positive-frequency overtones in an
    attempt to achieve greater accuracy in describing the waveform at earlier times using linear perturbation theory. 
    It has been shown that the inclusion of higher overtones can shift the linear regime to the peak of $(l,m)=(2,2)$ spherical harmonic mode. 
    In this work, we show that the inclusion of negative-frequency (counterrotating) modes, called `mirror' modes, extends the validity of linear perturbation 
    theory to even earlier times, with far lower systematic uncertainties in the model in recovering remnant parameters 
    at these early times.
    A good description of the signal at early times also enables for a greater signal-to-noise ratio to be accumulated in the 
    ringdown phase, thereby, allowing for a more accurate measurement of remnant parameters and tests of general relativity.
\end{abstract}

\keywords{Gravitational waves, binary black hole, ringdown, black hole perturbation theory}
\maketitle

\section{Introduction}

A perturbed black hole (BH) settles down to a stationary state by the emission of gravitational waves. 
At late times, when the perturbations are small and backreaction is not substantial, emitted gravitational waves 
form a discrete spectrum of complex frequencies called quasi-normal modes (QNMs) \citep{Vishveshwara:1970zz}, 
sometimes referred to as the ringdown signal. 
For a Kerr BH, QNMs are completely specified by its mass $M_f$ and dimensionless spin $a_f$.
This is a consequence of the `no-hair' theorem \citep{Misner:1974qy}. For a given angular mode $(l,m)$, there are 
a countably infinite number of QNMs ($\omega_{lmn} \in \mathbb{C}$) characterized by their overtone index ($n=0,1,2,...$).
The overtone numbers are assigned in decreasing order of damping times, i.e., the lowest overtone
number ($n=0$) has the largest damping time and is, therefore, the longest lived.
It then follows that, if one takes the starting time of the ringdown
to be at a late enough time after merger, then all
the higher overtones would have decayed sufficiently so that the ringdown signal can be 
described by a single overtone.

The information about the nature of the initial perturbation is contained in the complex excitation 
amplitude of each QNM. For a binary black hole (BBH) merger, then, the excitation amplitudes, in general, depend 
on the binary parameters like the mass ratio ($q$), the spin angular momenta of the two component BHs ($\vec{s_1}$, $\vec{s_2}$), 
and the eccentricities of the binary orbit ($e_1$, $e_2$). 

Historically, in gravitational wave data analysis, the start time for ringdown was choosen so  
that not only the non-linearities had died down but also the higher overtones had sufficient time to decay.
This made it possible for ringdown to be modeled using only the most dominant QNM \citep{FlanHugh98}. 
Recently, however, there have been efforts to model the ringdown signal using higher overtones \citep{PhysRevX.9.041060,Ota:2019bzl}
by starting the ringdown at earlier times when the contribution of the higher overtones to the ringdown signal is still significant. 
This has mostly been due to a three-fold reason. 

First, most of the BBH mergers observed by LIGO/Virgo~\citep{aLIGO_ref,AdV_ref} have nearly equal masses and small spins \citep{LIGOScientific:2018mvr}. For a non-spinning, equal
mass binary, the next dominant mode after $(l,m)=(2,2)$ 
is $(l,m)=(3,2)$ whose amplitude is a few percent compared to the dominant mode \citep{Kamaretsos:2011um,Borhanian:2019kxt}. 
For a consistency test of `no-hair' theorem, one determines the mass and spin of the perturbed BH using 
the dominant mode \citep{Echeverria:1989hg} and uses these estimates to determine the oscillation frequency and damping time of a 
subdominant mode. One then checks for its consistency with the measured value of the oscillation frequency and 
damping time of the subdominant mode \citep{BHspect04}.
For the current 
detector sensitivities and the BBH mergers we have observed so far, neither is this subdominant mode detectable 
nor is the frequency and damping time of the mode resolvable \citep{Carullo:2019flw,Brito:2018rfr} from the ringdown 
signal alone. 
Higher overtones are excited even for non-spinning, equal 
mass mergers and, therefore, measurement of the overtone frequencies and damping times
can be used for testing the `no-hair' theorem \citep{Isi:2019aib}.

Secondly, including overtones in a ringdown model can shift the start time of ringdown to earlier times
and can therefore increase the signal-to-noise (SNR) contained in the ringdown.\footnote{\citet{Bhagwat:2019dtm} 
showed that the SNR may not always increase on the addition of higher overtones depending on the 
relative phase of the different overtones.} Indeed, \citet{PhysRevX.9.041060} showed that including 
up to $n=7$ overtones can shift the start time of ringdown to the peak of $h_{22}$ mode. 

Finally, LISA could observe BBH mergers with total mass greater than $10^8M_{\odot}$, 
which would have a very small or no inspiral part \citep{Kamaretsos:2011um}.
Having a ringdown model where multiple excitation amplitudes have been mapped to progenitor
parameters can, then, be used to determine the parameters of the binaries.

There have been numerous studies in literature that model the ringdown phase of a BBH merger signal using higher angular 
modes \citep{Kamaretsos:2011um,London:2014cma,Baibhav:2017jhs,London:2018gaq}.  
Other studies model the ringdown phase using higher overtones \citep{PhysRevX.9.041060,Ota:2019bzl}.
\citet{Cook:2020otn} does a multimode fitting of the ringdown phase including higher overtones.
The effective one-body (EOB) formalism has modeled ringdown using a superposition of QNMs and 
psudomodes (modes that are not QNMs) \citep{BCP07NR,Pan:2013rra,Taracchini:2013rva,Babak:2016tgq}. 
For a discussion on some of these studies, see \citet{PhysRevX.9.041060}. 

In all of these studies, the ringdown is modeled using only the positive-frequency (corotating) part
of the QNM spectrum.\footnote{\citet{Taracchini:2013rva} includes some negative-frequency modes together with smoothing
functions to provide a smooth transition from inspiral to merger-ringdown.} 
\citet{London:2014cma} do look for negative-frequency modes using their \textit{greedy-OLS}
algorithm but do not find them to be significantly excited for non-spinning binaries.
\citet{Forteza:2020hbw} fit negative-frequency modes to a BBH merger signal for an $n=1$ overtone model and find that the lower-order
counterrotating modes are not significantly excited.
\citet{Hughes:2019zmt} and \citet{PhysRevD.100.084032} numerically solve the Teukolsky equation for a 
point particle infall into a Kerr BH and find negative-frequency modes are excited.

We will refer to the negative-frequency modes as `mirror' modes and positive-frequency modes 
as `regular' modes from hereon.
For a BBH merger there is no reason, apriori, for the gravitational waves to consist of
regular modes alone (see \citet{PhysRevD.73.064030} for more discussion). 
For the mirror modes to be omitted from ringdown waveforms consistently 
(especially ones including higher overtones), it has
to be shown that the excitation amplitudes for these modes are much
smaller than regular modes. 
Alternatively, one can argue that these mirror modes start at an earlier time than the regular modes
and owing to their smaller damping times than their corresponding regular modes, 
they decay away before the ringdown starts for the (dominant) regular modes.

In this paper, we study the effect of including mirror modes in a gravitational
waveform. 
We fit the complex excitation amplitudes to numerical relativity (NR) waveforms and
show that including mirror modes in a ringdown waveform improves the
fits to NR waveforms at all times in the ringdown regime. 
The improvement in fits is especially
prominent at times before the peak of the $(l,m)=(2,2)$ spherical harmonic mode. 
We study the systematics of the modeling to determine if the improvement in the fits
to NR waveforms is due to the presence of mirror modes in the gravitational waveform.
An alternative reasoning for the enhancement of the fits could be that the additional free parameters introduced in the model
due to the inclusion of mirror modes acts as basis functions and fit to some of the 
non-linearities in the waveform, especially at early times. 

We note that most of the metrics used in this study have been introduced in
\citet{PhysRevX.9.041060} and \citet{Bhagwat:2019dtm} to study the importance of including higher (regular) 
overtones to a ringdown model. 
We will use their model as a reference and compare the results of our model 
against theirs. 

Our goal is to improve the theoretical modeling of ringdown waveforms. 
We make the case that including mirror modes in a ringdown model give better estimates of the remnant parameters
at times before the peak of the $(l,m)=(2,2)$ spherical harmonic mode.
We point out that the detectability and resolvability of mirror modes is beyond
the scope of this paper (see \citet{Bhagwat:2019dtm,Cabero:2019zyt,Isi:2019aib} for a discussion
on detecting higher angular modes and overtones).
The start time of ringdown has also been a contentious topic in the literature and we refer the interested reader to 
\citet{Kamaretsos:2011um,Nollert:1999ji,Berti:2007fi,Baibhav:2017jhs,Carullo:2018sfu} for a discussion on the 
different choices that have been made in the literature.

The paper is organised as follows. In section \ref{sec:BBHwaveform} we introduce the 
ringdown model and lay down the assumptions and approximations used. In section \ref{sec:results}
we show our results and discuss modeling systematics. 
In section \ref{sec:conclusion} we conclude by highlighting the main results of the paper 
and discuss some issues with a pure ringdown model.

\section{Binary black hole ringdown waveform}
\label{sec:BBHwaveform}

The gravitational waves emitted by a perturbed Kerr BH of mass $M_f$ and spin $a_f$ 
as observed by an observer at a large distance $r$
is given by\footnote{We use the sign convention used in SXS simulations.} \citep{Press:1973zz},
\begin{equation}
    h(t) = h_{+} - i h_{\times} = \frac{M_f}{r}\sum_{lmn}\mathcal{C}_{lmn} e^{-i \omega_{lmn}t} \prescript{}{-2}S_{lm}(\Omega,a_f\omega_{lmn}) ,
\end{equation}
where $\mathcal{C}_{lmn}=\mathcal{A}_{lmn}e^{-i \phi_{lmn}}$ are the complex excitation amplitudes, $t$ is the retarted time at null infinity, 
and $\Omega$ is the coordinate on a sphere $(\theta,\phi)$. 
The complex $\omega_{lmn} \equiv \omega_{lmn}(M_f,a_f)$ are the QNM frequencies as determined by perturbation theory. 
The complex angular functions $\prescript{}{-2}S_{lm}(\Omega,a_f\omega_{lmn})$ are the $-2$ spin-weighted spheroidal harmonic functions 
under which the perturbation equations of a Kerr BH decompose into a radial and angular part. 
They reduce to $-2$ spin-weighted spherical harmonic functions $\prescript{}{-2}Y_{lm}(\Omega)$ in the 
Schwarzschild BH case ($a_f=0$). 

\begin{figure}[h]
    \centering
    \includegraphics[width=\columnwidth]{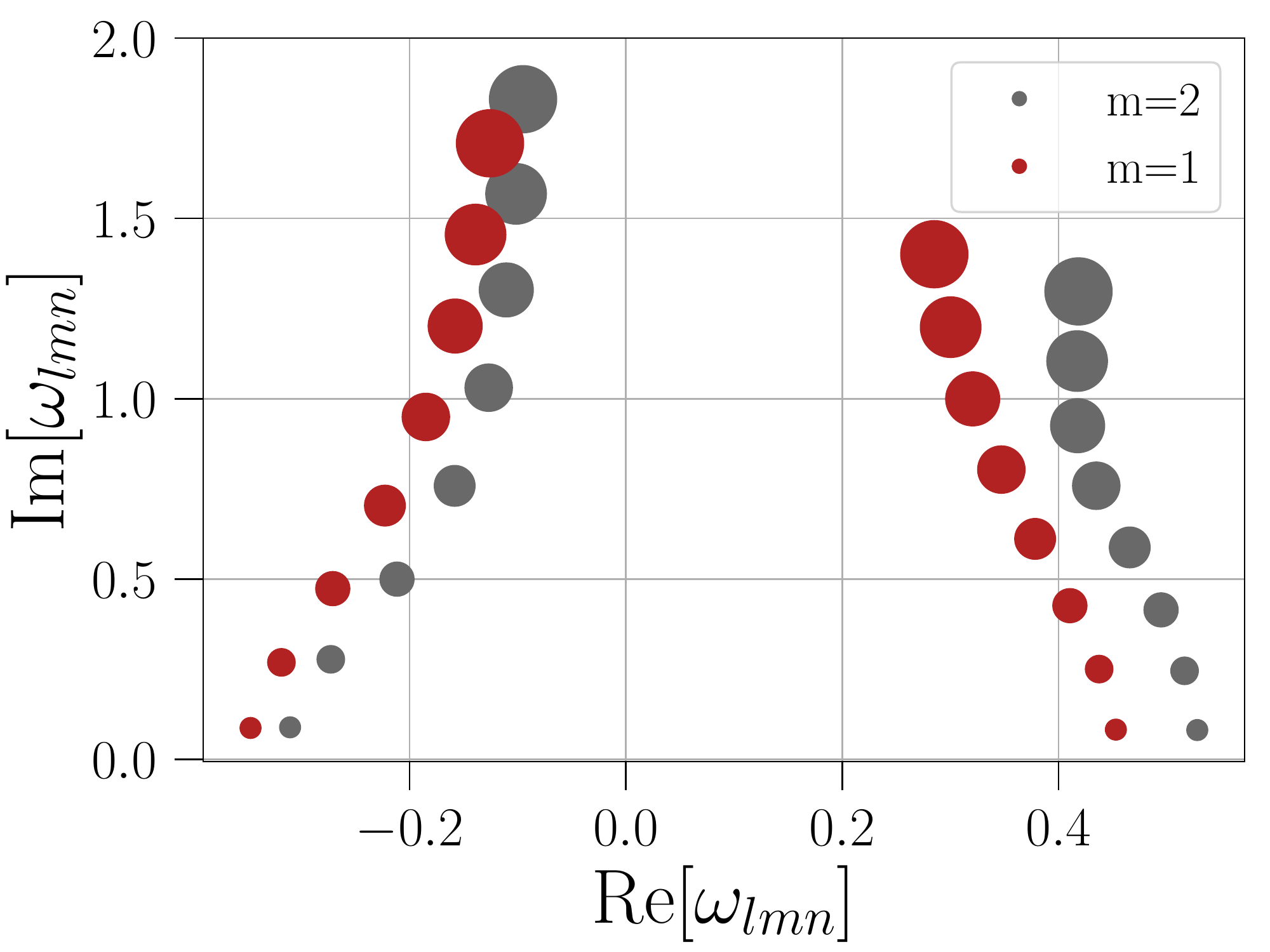}
    \caption{QNM frequencies of the first 8 overtones ($n=0, 1, 2,..., 7$) for $l=2$ and $m>0$ for a Kerr BH 
        with dimensionless spin $a_f=0.69M$ with larger flecks denoting higher overtones.  
        Notice that the damping time of the negative-oscillation frequency solution is always shorter than its 
        corresponding positive-oscillation frequency solution. The QNM frequencies for $m<0$ can be easily
        worked out using Eq.~(\ref{eq:kerr_sym}).}
    \label{fig:qnm}
\end{figure}

The perturbation of a Kerr BH is described by the Teukolsky equation \citep{Teukolsky:1973ha}.
The Teukolsky equation is a second order differential equation and, therefore,
for a given angular mode $(l,m)$ and overtone number $n$, there are two linearly independent solutions. 
For a perturbed Schwarzschild black hole, due to the spherical symmetry of the background, if 
$\omega_{lmn}$ is one of the solutions, the second linearly independent solution is given by $-\omega_{lmn}^*$, 
i.e., positive- and negative-oscillation frequency solutions have the same damping time.
For a Kerr black hole there is no such simple relationship between the two solutions because of the reduced
symmetry of the system. 
One still has positive- and negative-oscillation frequency solutions, though, 
with the damping times of the positive oscilaltion frequency solution always larger than that of the corresponding negative
oscillation frequency solution (see Fig.~\ref{fig:qnm} for an example). 
The azimuthal symmetry of the Kerr background does, however, separate the angular part of the 
perturbation equations in terms of $-2$ spin-weighted spheroidal harmonics,
with the radial equations obeying the following symmetry relations:
\begin{equation}
    \label{eq:kerr_sym}
    \omega_{lmn} = -\omega_{l-mn}^{*} \,, \;\;\;\;\;\; A_{lmn} = A_{l-mn}^{*} 
\end{equation}
where $A_{lmn}$ are the angular separation constants.\footnote{Not to be confused with $\mathcal{A}_{lmn}$
which are the real-valued excitation amplitudes.}

A gravitational waveform, in general, is therefore a linear combination of the two solutions and is
given by,\footnote{The calculation in the remainder of the section follows closely that of 
\citet{PhysRevD.73.064030} and \citet{PhysRevD.100.084032}.}
\begin{equation}
    \begin{split}
    \label{eq:PT_waveform0}
        h(t) = h_{+} - i h_{\times} =& \frac{M_f}{r}\sum_{lmn}[\mathcal{C}_{lmn} e^{-i \omega_{lmn}t} \prescript{}{-2}S_{lm}(\Omega,a_f\omega_{lmn}) \\
        &+ \mathcal{C}_{lmn}' e^{-i \omega_{lmn}'t} \prescript{}{-2}S_{lm}'(\Omega',a_f\omega_{lmn})] \\
    \end{split}
\end{equation}
where $\Omega'=(\pi-\theta,\phi)$.

Equation (\ref{eq:PT_waveform0}) can be simplified using the symmetry relation Eq. (\ref{eq:kerr_sym}) to give
\begin{equation}
    \begin{split}
    \label{eq:PT_waveform}
        h(t) = h_{+} - i h_{\times} =& \frac{M_f}{r}\sum_{lmn}[\mathcal{C}_{lmn} e^{-i \omega_{lmn}t} \prescript{}{-2}S_{lm}(\Omega,a_f\omega_{lmn}) \\
        &+ \mathcal{C}_{lmn}' e^{i \omega_{l-mn}^{*}t} \prescript{}{-2}S_{l-m}^{*}(\Omega',a_f\omega_{lmn})]
    \end{split}
\end{equation}

Numerical relativity simulations, in general, decompose the angular part of the waveform in spherical 
harmonic functions and the ringdown part can be written as
\begin{equation}
    \label{eq:NR_waveform}
    h(t) = h_{+} - i h_{\times} = \frac{M_f}{r} \sum_{lm} h_{lm}(t) \prescript{}{-2}Y_{lm}(\Omega)
\end{equation}
In order to comapre an NR waveform to a perturbation theory ringdown waveform, we have to expand the
spheroidal harmonic functions in a basis of spherical hamonics.

The $-2$  spin weighted spheroidal functions can be expressed in an orthonormal basis of $-2$ spin-weighted 
spherical harmonics as 
\begin{equation}
    \prescript{}{-2}S_{lm}(\Omega;a_f\omega_{lmn}) = \sum_{l'm'} \mu_{lm}^{l'm'}(a_f\omega_{lmn}) \prescript{}{-2}Y_{l'm'}(\Omega)
\end{equation}
where $\mu_{lm}^{l'm'} = \mu_{lm}^{l'} \delta_{m}^{m'}$.

Equating the left hand side of Eq. (\ref{eq:PT_waveform}) and (\ref{eq:NR_waveform}) and using the 
orthogonality condition for spin-weighted spherical harmonics, we can write the gravitational 
waveform in terms of spherical harmonics as
\begin{equation}
    \begin{split}
        h(t) = h_{+} - i h_{\times} =& \frac{M_f}{r} \sum_{l'm}\sum_{ln} [\mathcal{C}_{lmn}\mu_{ln}^{l'} e^{-i \omega_{lmn}t} \\
        &+ (-1)^l\mathcal{C'}_{lmn}\mu_{ln}^{*l'} e^{i \omega_{l-mn}^{*}t}] \prescript{}{-2}Y_{l'm}(\Omega)
    \end{split}
\end{equation}

An NR angular mode is then related to the excitation amplitudes by
\begin{equation}
    \label{eq:hlm}
    h_{l'm} = \sum_{ln} [\mathcal{C}_{lmn}\mu_{ln}^{l'} e^{-i \omega_{lmn}t} + (-1)^l\mathcal{C'}_{lmn}\mu_{ln}^{*l'} e^{i \omega_{l-mn}^{*}t}]
\end{equation}
where we have used the following relation
\begin{equation}
    \prescript{}{-2}Y_{lm}^{*}(\pi-\theta,\phi) = (-1)^l \prescript{}{-2}Y_{l-m}(\theta,\phi)
\end{equation}

Note that a spherical harmonic mode ($l',m$) has contributions from spheroidal harmonic orbital angular 
quantum numbers $l$ other than the corresponding spherical harmonic one, $l'$. 
In this study we focus on $l'=2$ mode of a nearly equal mass
binary (SXS:BBH:0305) and a high mass ratio binary (SXS:BBH:1107).
In both the cases the higher $(l',m)$ modes are subdominant and contribute at a sub-percent
level to the (2,2) mode once spherical-spheroidal mixing coefficients are taken into account \citep{PhysRevX.9.041060,Borhanian:2019kxt}. 
The ringdown model, therefore, simplifies to 
\begin{equation}
    h_{lm}(t) = \sum_{n} [\mathcal{C}_{lmn} e^{-i \omega_{lmn}t} + \mathcal{C'}_{lmn} e^{i \omega_{l-mn}^{*}t}]
\end{equation}
where the $(-1)^l$ term in Eq. (\ref{eq:hlm}) has been absorbed in to the arbitrary coefficients $\mathcal{C'}_{lmn}$.

\section{Results}
\label{sec:results}

In the previous section we introduced our ringdown model and elucidated the assumptions that were made.
In this section we show that our model agrees with NR better than the reference model. 
We further show that the errors in the estimation of remnant parameters is smaller for our model 
at times before $h_{22}$ mode peaks. 

The complex amplitudes in the ringdown waveform is a function of the effective potential of a BH spacetime
and initial condition for perturbations. It is a highly non-trivial initial value problem. 
Analytic solutions exist for special cases of point particles falling into a BH \citep{Sun:1988tz,Berti:2006wq,Hadar:2009ip,Zhang:2013ksa}.
Therefore, for a binary black hole ringdown waveform, the excitation amplitudes have to be inferred
by fitting a ringdown waveform to NR simulations. 

\begin{figure}[h]
    \centering
    \includegraphics[width=\columnwidth]{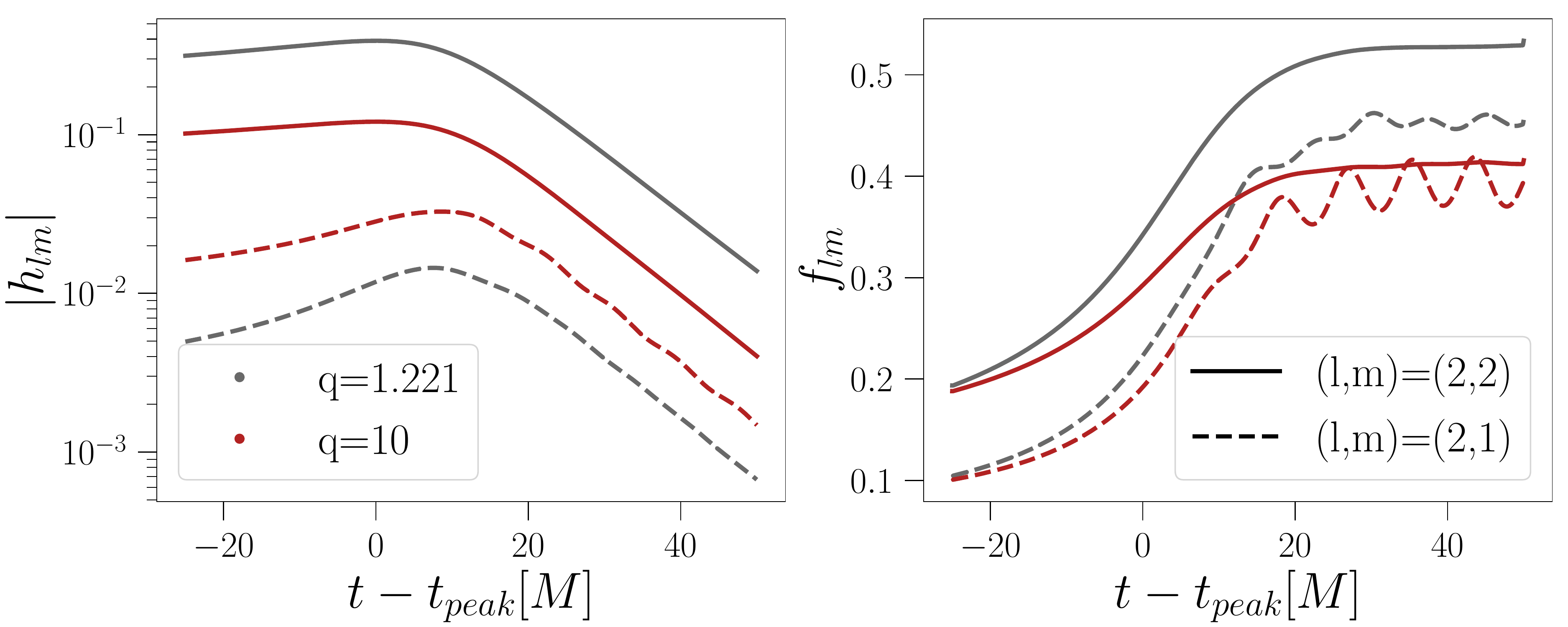}
    \caption{The amplitude (left) and mode frequency (right) evolution of the $l=2$ modes for an almost equal mass ($q=1.221$) 
        binary (grey) and a large mass ratio ($q=10$) binary (red). 
        Solid (dashed) lines show the $m=2$ ($m=1$) mode.}
    \label{fig:waveform}
\end{figure}

A second point of note is that, due to the spin of a Kerr BH,
even if the initial conditions have a definite mode structure, both corotating and counterrotating (mirror) modes
will be excited in response to the initial perturbations~\citep{Dorband:2006gg,Nagar:2006xv,Damour:2007xr,Bernuzzi:2010ty,Zimmerman:2011dx}.
In Fig.~\ref{fig:waveform}, we plot the evolution of the amplitude $|h_{lm}|$ and the mode frequency 
$f_{lm}=-\Im(\dot{h}_{lm}/h_{lm})$
of the spherical harmonic modes $l=2$ as a function of the retarded time for the two cases under study.
For the $m=1$ mode, in both the cases, the modulation of the amplitude and the mode frequency due to the fundamental
mirror mode can be distinctly seen in the figure. 
For the $m=2$ mode, the modulations are not visible to the eye. 
Clearly, the excitation amplitudes of the mirror modes depend on the value of the azimuthal quantum number $m$~\citep{Damour:2007xr}.

We consider two test cases from the publicly available SXS\footnote{Simulating eXtreme Spacetimes \citep{SXS}} catalogue of NR simulations,
SXS:BBH:0305 and SXS:BBH:1107, corresponding to non-spinning binaries with mass ratios $q=1.221$ and
$q=10$, respectively.
The former is a GW150914-like signal with the final mass $M_f=0.9520 M$ and dimensionless spin $a_f=0.6921$.
The later has a final mass $M_f=0.9917 M$ and spin $a_f=0.2605$.
The QNM frequencies are fixed to their GR values (and calculated using Ref. \citep{Stein:2019mop}) which leaves only the complex amplitudes
as free parameters which we fit to NR using a linear least squares method. 
A ringdown model with overtone index upto $N$ has $2(N+1)$ complex amplitudes that are being fit.\footnote{
\citet{Bhagwat:2019dtm} calls a ringdown model with $N$ included overtones an ($N+1$)-tone model and we will follow their nomenclature.}
We vary the start time of ringdown from $t_0 = t_{\rm peak} -25M$ to $t_0 = t_{\rm peak} + 50M$, where the origin $t_0 = t_{\rm peak}$ has been taken to 
be the peak of the $h_{22}$ mode. 
We fix the end time to be $t = t_{\rm peak} + 90M$ by which time even the longest lived overtone would have decayed
essentially to numerical noise.
We define the mismatch between the best-fit ringdown waveform ($h_{\rm fit}$) and NR waveform ($h_{\rm NR}$) by
\begin{equation}
    \mathcal{M} = 1 - \frac{\langle h_{\rm NR}|h_{\rm fit} \rangle}{\sqrt{\langle h_{\rm NR}|h_{\rm NR} \rangle \langle h_{\rm fit}|h_{\rm fit} \rangle}},
\end{equation}
where the inner product is defined as 
\begin{equation}
    \label{eq:ip}
    \langle a(t)|b(t) \rangle = \int_{t_0}^{90M} a^{*}(t)b(t) dt .
\end{equation}

We note that QNMs are not orthogonal and complete under this inner product.
This has been a longstanding theoretical question and it is doubtful whether such an 
inner product can be defined for QNMs that is also of practical use \citep{Nollert:1998ys}. 

\begin{figure*}[h]
    \centering
    \subfloat{
        \includegraphics[width=\columnwidth]{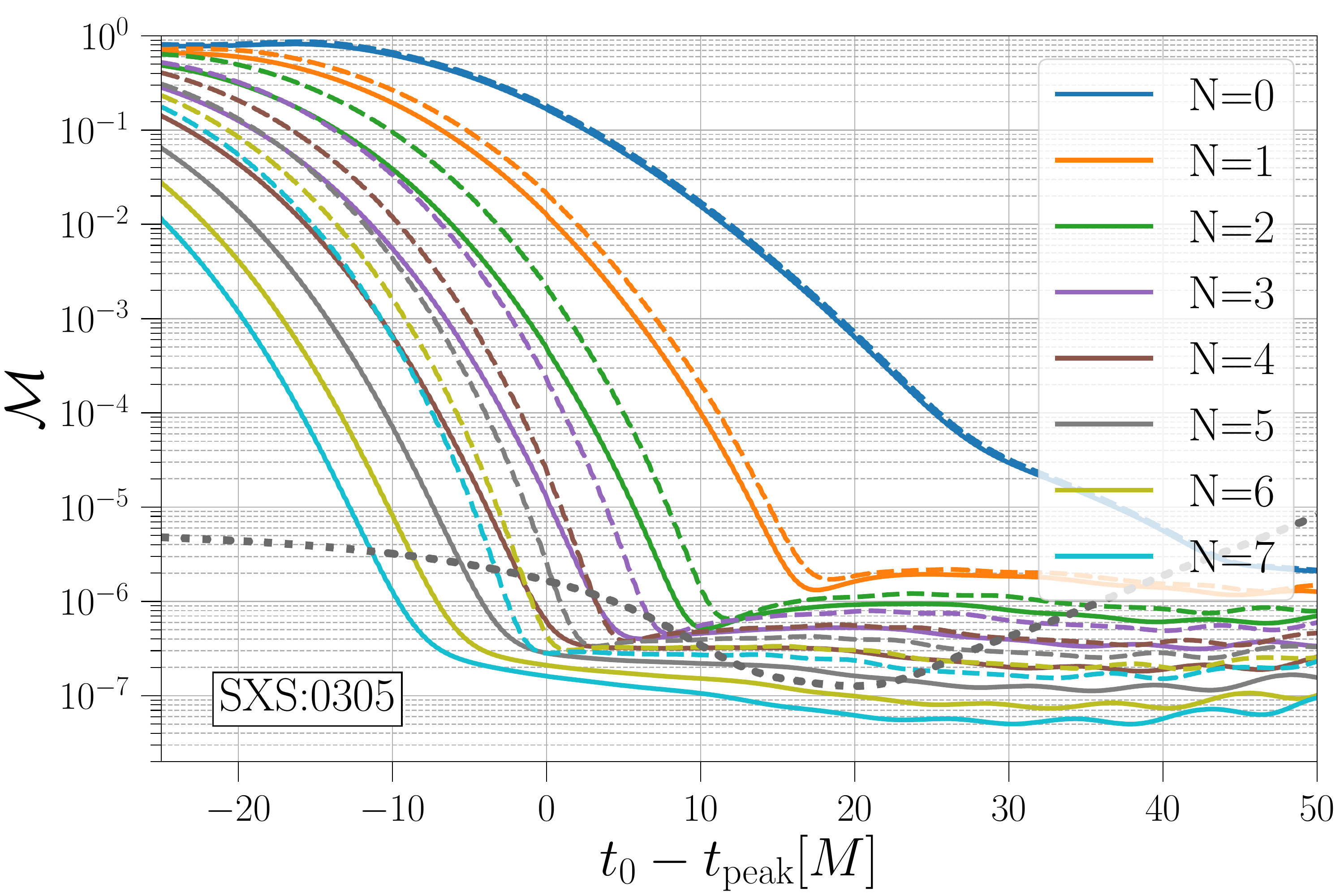}
    }
    \subfloat{
        \includegraphics[width=\columnwidth]{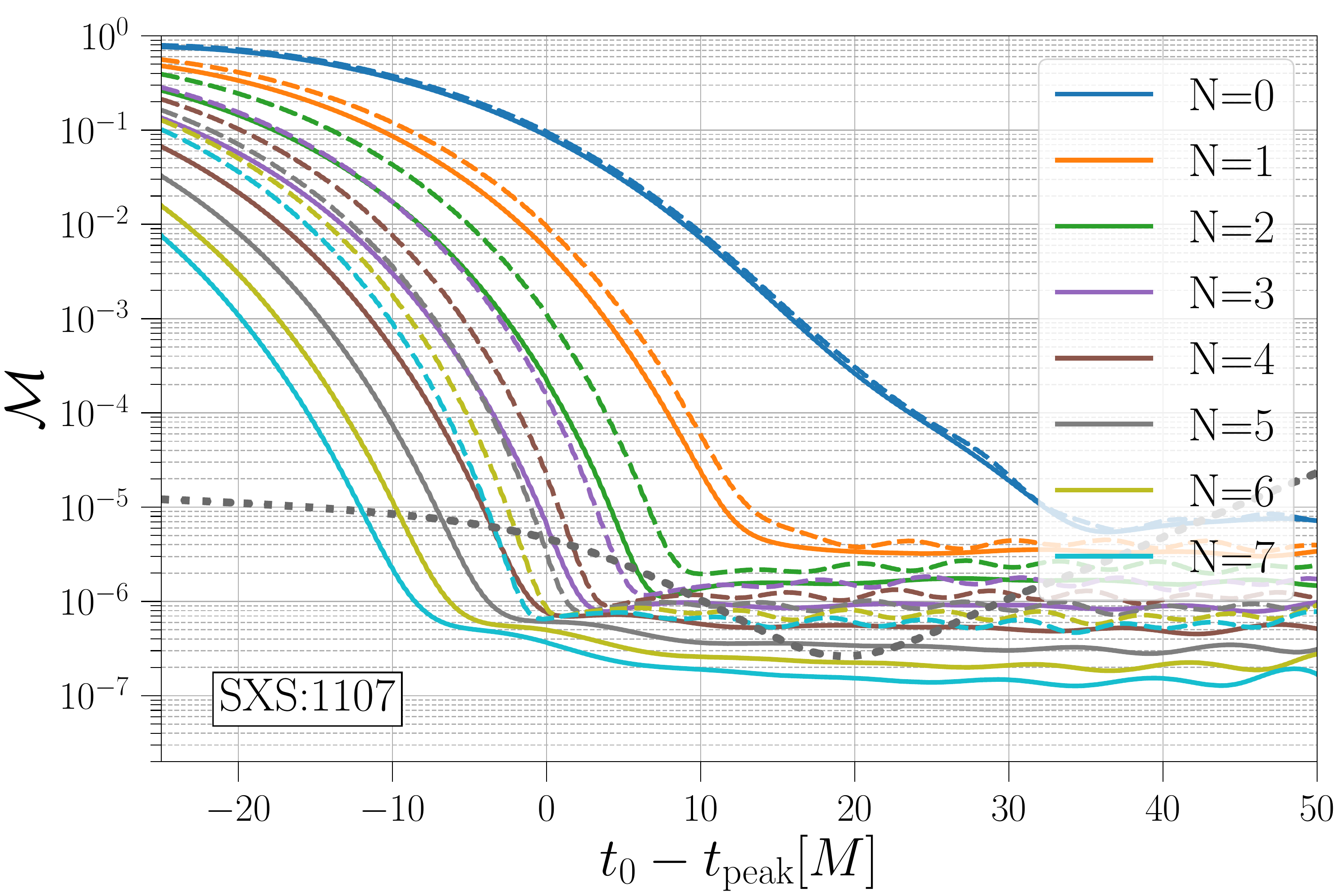}
    }

    \subfloat{
        \includegraphics[width=\columnwidth]{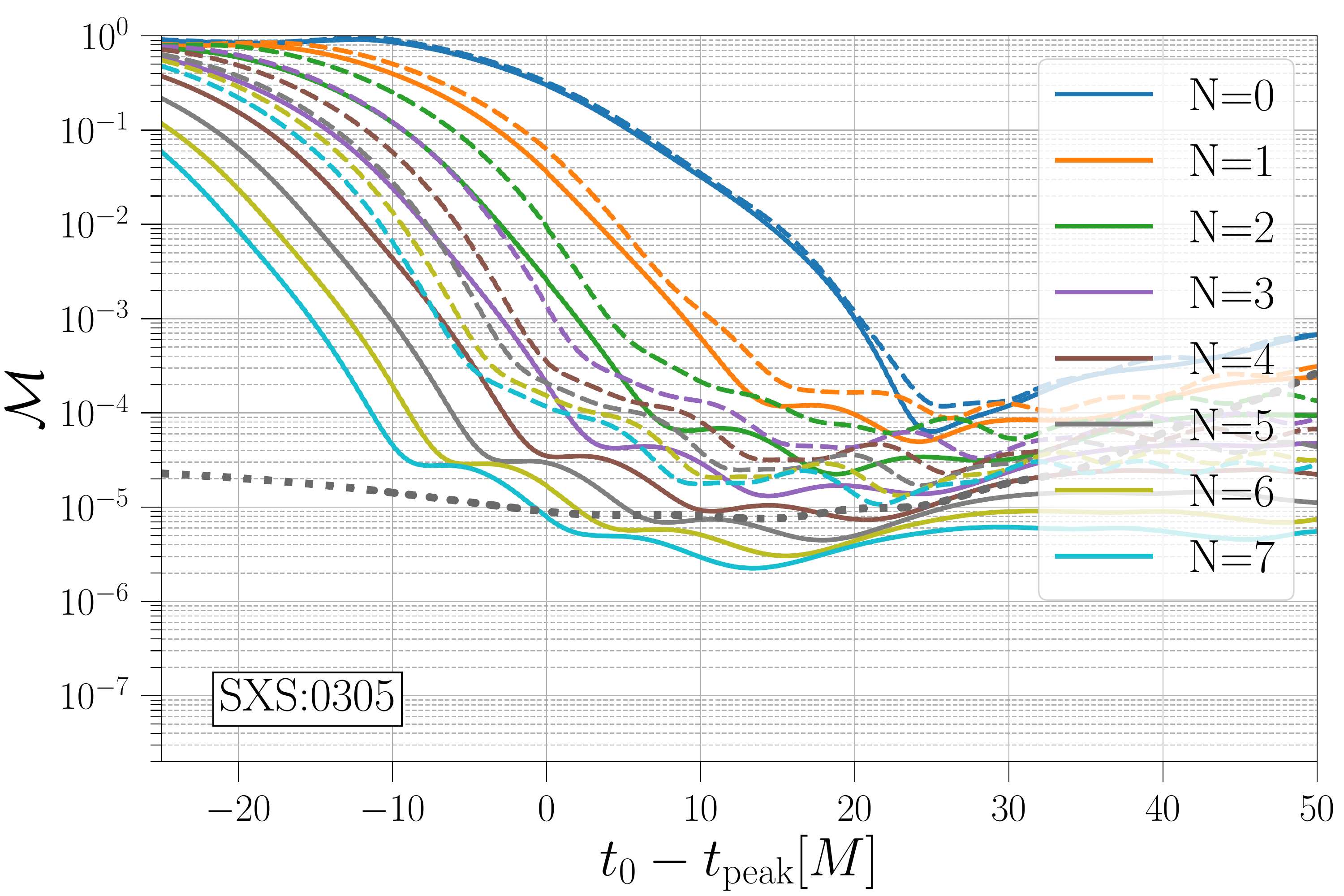}
    }
    \subfloat{
        \includegraphics[width=\columnwidth]{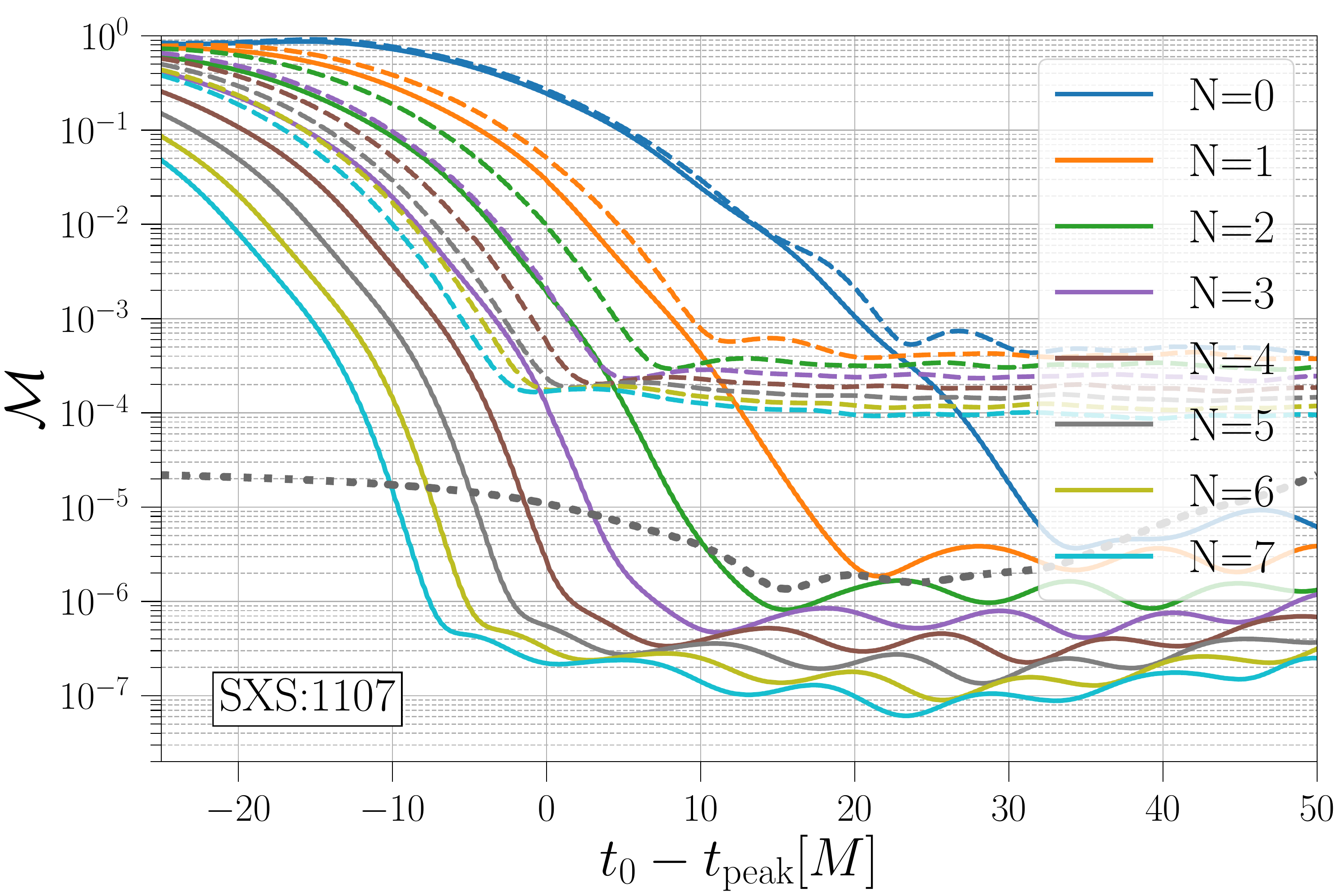}
    }
    \caption{Comparison of mismatches $\mathcal{M}$ between mirror modes model
        and our reference model for an $(N+1)$-tone model for the two cases under consideration. 
        The mismatches $\mathcal{M}$ are plotted 
        as a function of start time $t_0$ whose origin is taken to be the peak of $h_{22}$. 
        Solid (dashed) lines represent the class of ringdown models with (without) the mirror modes.
        The top panel shows the mismatches for $(l,m)=(2,2)$ spherical harmonic mode while the bottom panel depicts
        the same for $(l,m)=(2,1)$ mode.
        The grey, dotted, horizontal line shows the mismatch for the two highest resolution NR simulations.
        We note that the mismatches are always lower when the mirror modes are included
        with marked improvements at times before the peak of the $h_{22}$ mode.}
    \label{fig:mismatch}
\end{figure*}

In Fig. \ref{fig:mismatch}, we show the dependence of $\mathcal{M}$ as a function of the start time $t_0$.
We compare $\mathcal{M}$ between our model and the reference model of \citet{PhysRevX.9.041060} 
for up to an 8-tone ringdown waveform.
The mismatch curves are qualitatively similar for both the simulations apart from the characteristic
that the mismatches for the $(l,m)=(2,1)$ mode of the (almost) equal mass binary has multiple crests
and troughs before hitting the numerical noise floor.
We see that the mismatch for any given $t_0$ and ($N+1$)-tone model is lower for mirror mode
model compared to the reference model. The betterment in $\mathcal{M}$ is roughly 3 orders of magnitude
for an 8-tone model at $t_0 \sim -10M$.
We observe that the fundamental mirror mode is definitively excited in the $q=10$ binary and 
is considerable at late times as was seen in Fig.~\ref{fig:waveform}.
We note that the mismatches for $(l,m)=(2,2)$ mode of SXS:BBH:0305 agree with Fig. 1 of \citet{PhysRevX.9.041060} when mirror modes are not included,
thus providing a validation for our fits.

In addition, in Fig.~\ref{fig:mismatch}, we compute the numerical noise floor by calculating the
mismatch between the waveforms of the two highest resolved NR simulations.
This would give us an estimate of the truncation errors in an NR simulation due to finite grid sizes. 
For the modes and the cases under study, we see that the mismatch of this noise floor is
between $\sim10^{-4}$ and $\sim10^{-7}$ with the higher mismatches occuring at late times.
This is illustrated by the grey horizontal dotted lines in the figure.

The plot shows another crucial feature. An 8-tone ringdown model
gives orders of magnitude lower mismatches than (say) a 2-tone model even when the ringdown is started
at a late enough time ($\sim t_{\rm peak}+30M$) when all higher overtones are expected to have decayed to numerical noise. 
This happens at or below the numerical noise floor of the simulations used and, therefore,
we believe that at these times the free parameters of the models are being fit to numerical noise 
and that this feature is unphysical. 

We point out that even though there is a huge improvement in $\mathcal{M}$ at early times with the 
inclusion of mirror modes in the ringdown model, mirror modes alone do not produce good fits. 
The positive oscillation frequency modes are still the dominant modes (at least for the cases considered in this 
study) present in the waveform. 
It would be interesting to see if spins and precession of the progenitor binary change this conclusion.

\begin{figure*}
    \centering
    \subfloat{
        \includegraphics[width=0.5\textwidth]{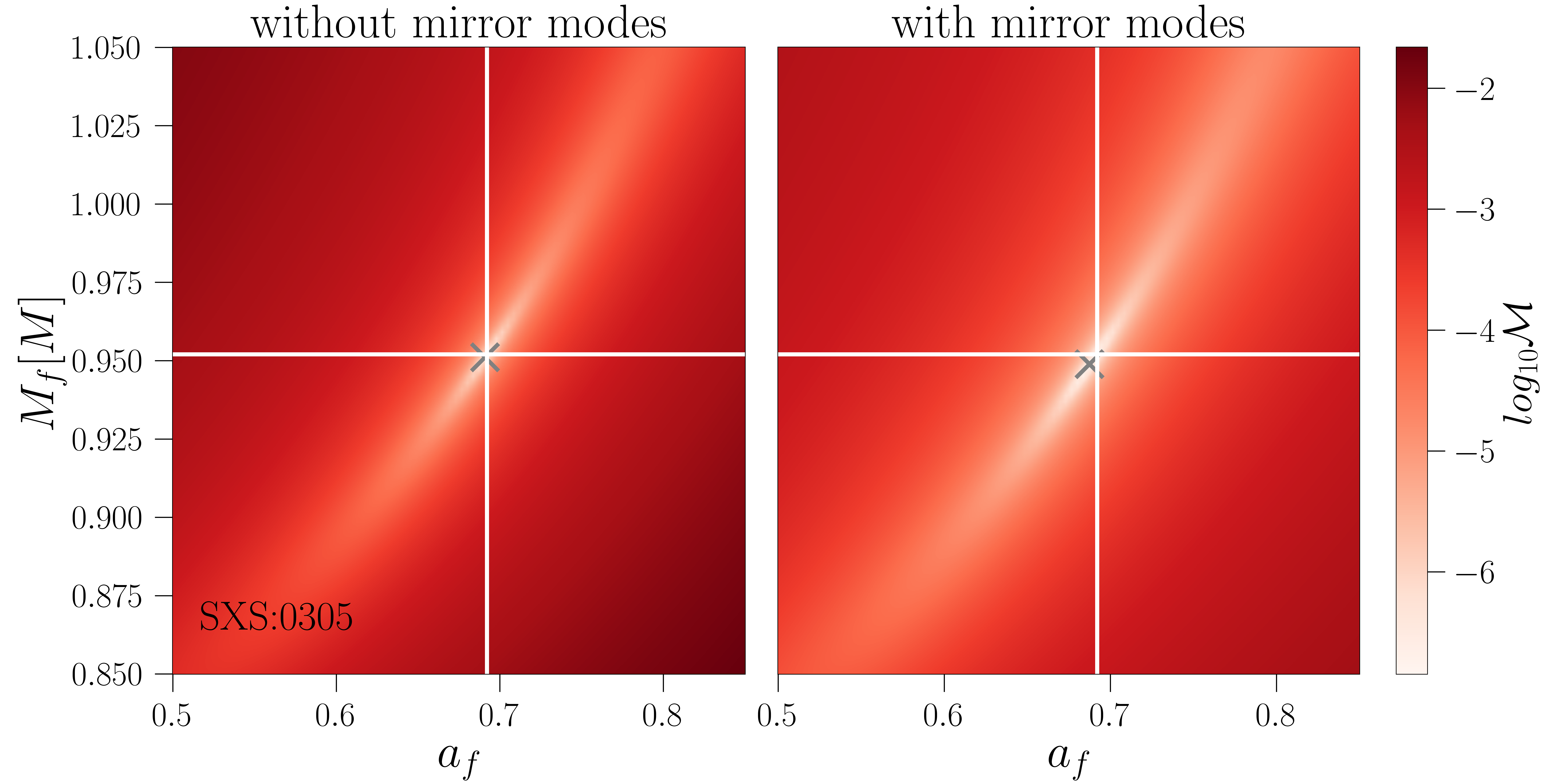}
    }
    \subfloat{
        \includegraphics[width=0.5\textwidth]{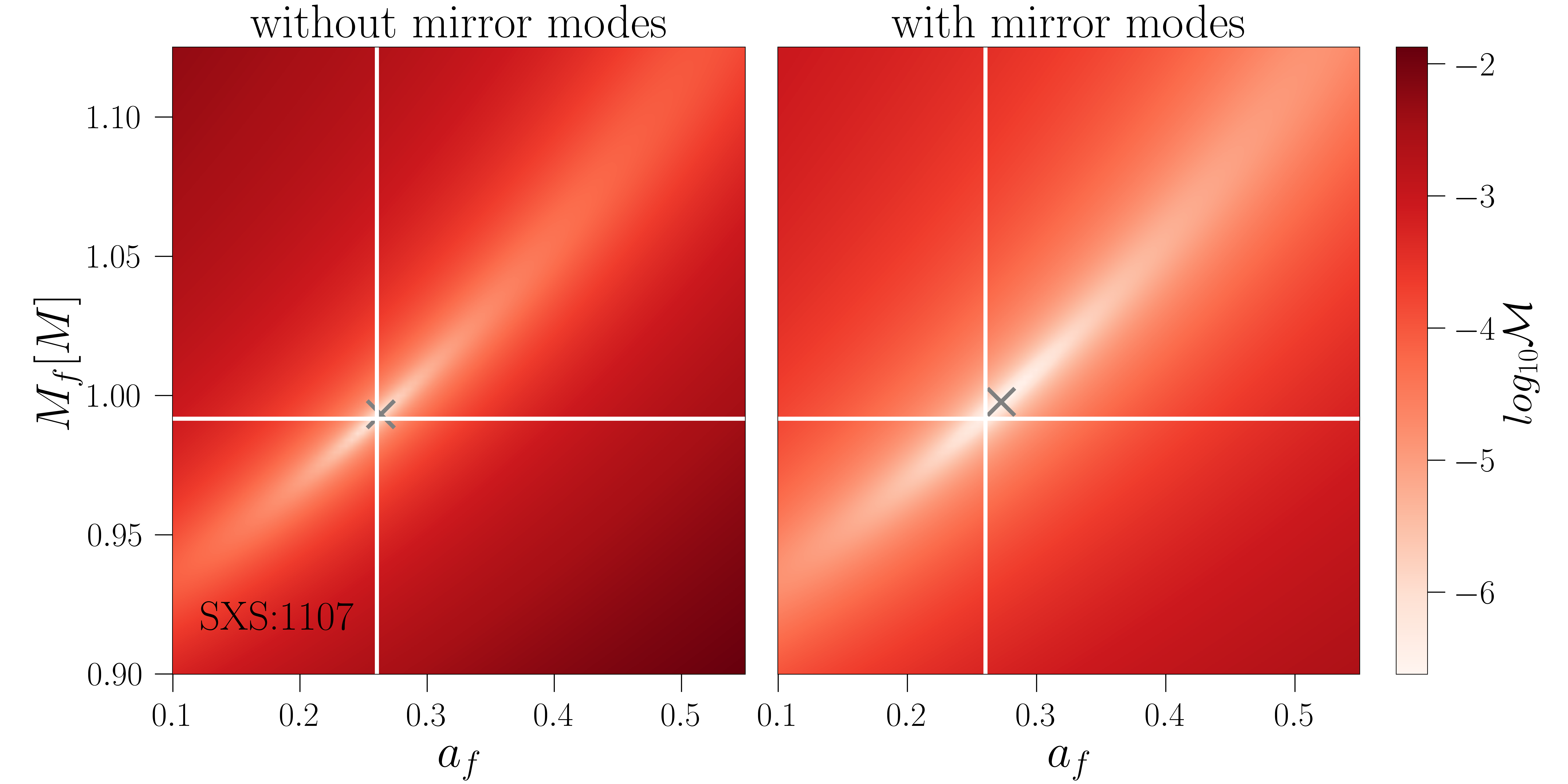}
    }

    \subfloat{
        \includegraphics[width=0.5\textwidth]{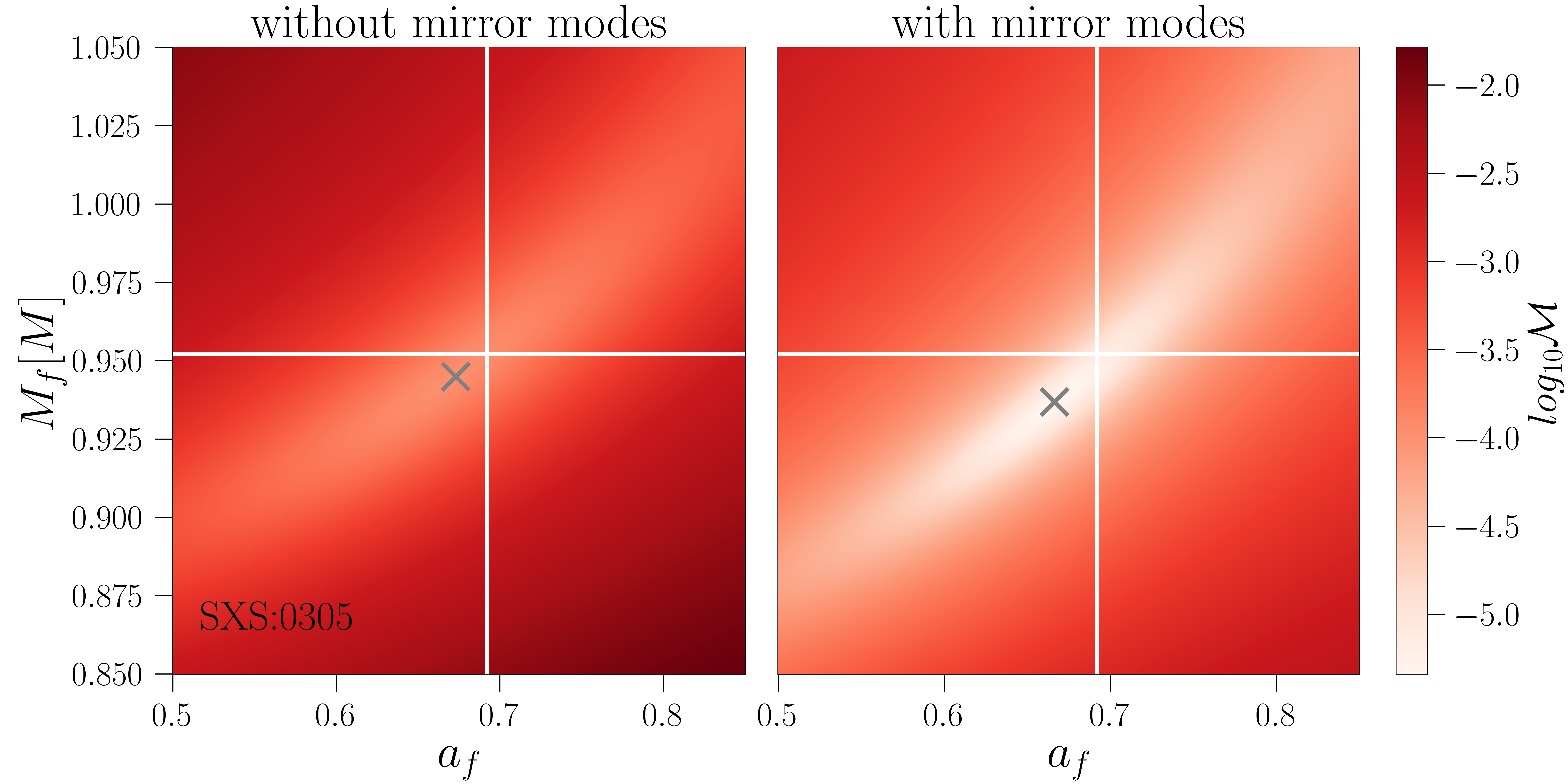}
    }
    \subfloat{
        \includegraphics[width=0.5\textwidth]{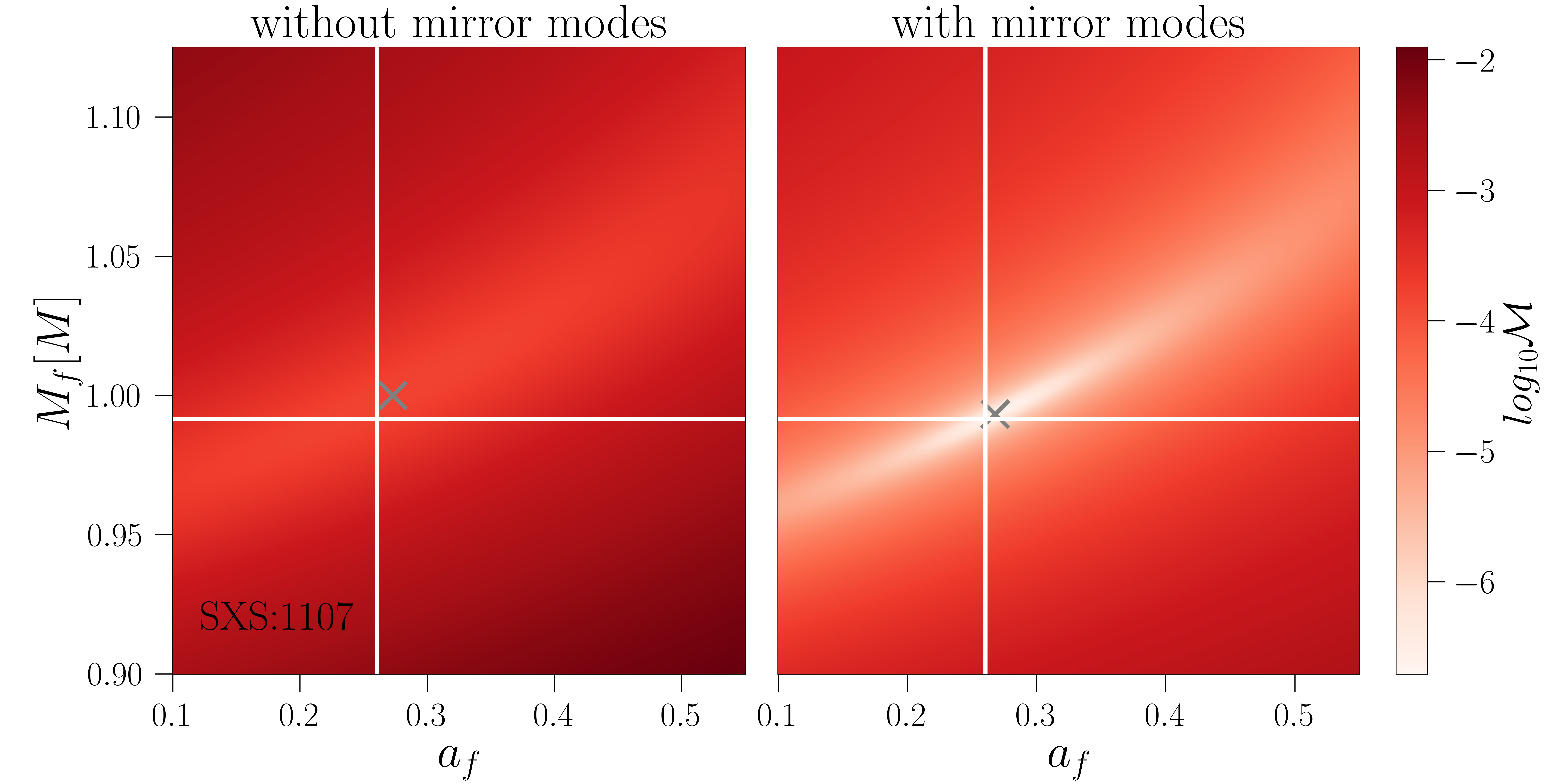}
    }
    \caption{Mismatch $\mathcal{M}$ between best-fit ringdown model and NR waveform for an 8-tone model
        on a grid of $M_f-a_f$ for GR QNM frequencies. The start time of ringdown is taken to be $t_0=0$.
        The white crosshair denotes the true remnant mass and spin as ascertained by NR while the grey cross
        shows the best-fit value determined by a ringdown model. 
        The result for the reference model is plotted on the left panel while the mirror modes model 
        is on the right panel of each plot.
        The result for spherical harmonic mode $(l,m)=(2,2)$ is plotted in the top panel of the figure while 
        the bottom panel highlights the result for the $l=2$, $m=1$ mode.
        The plot on the left is for the simulation SXS:0305 and the right plot is for SXS:1107.}
    \label{fig:mismatch_grid_0}
\end{figure*}

\begin{figure*}
    \centering
    \subfloat{
        \includegraphics[width=0.5\textwidth]{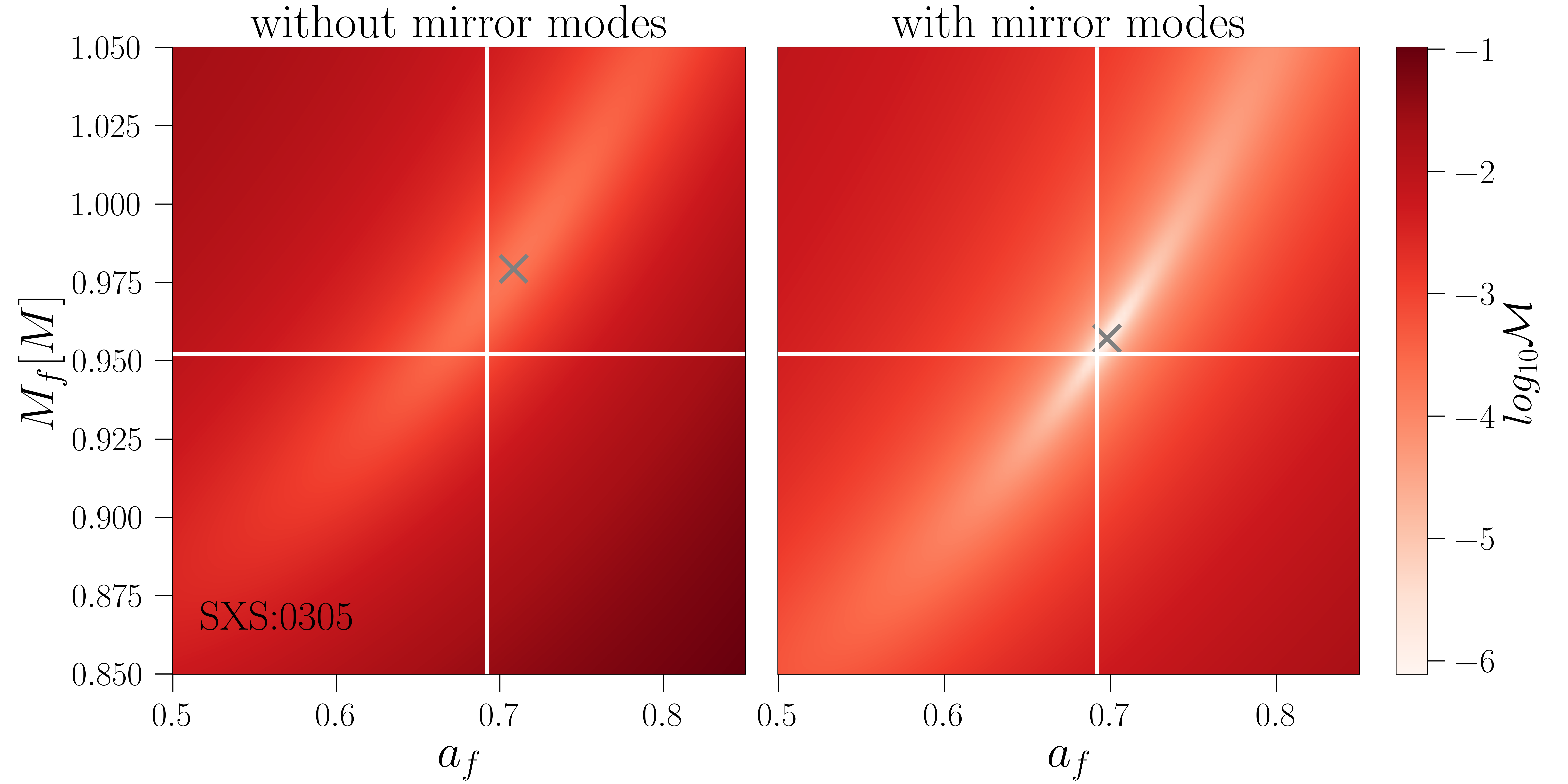}
    }
    \subfloat{
        \includegraphics[width=0.5\textwidth]{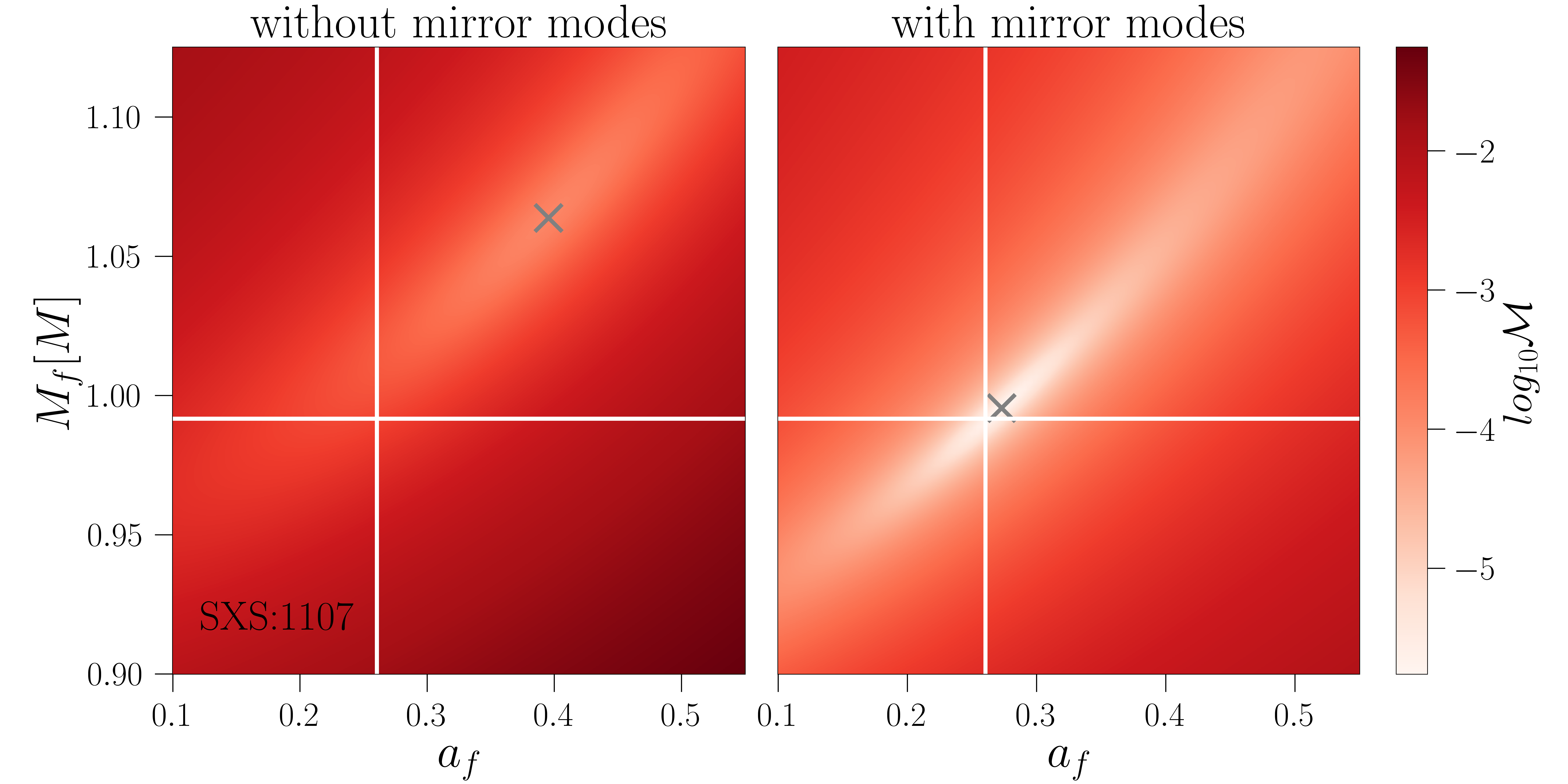}
    }

    \subfloat{
        \includegraphics[width=0.5\textwidth]{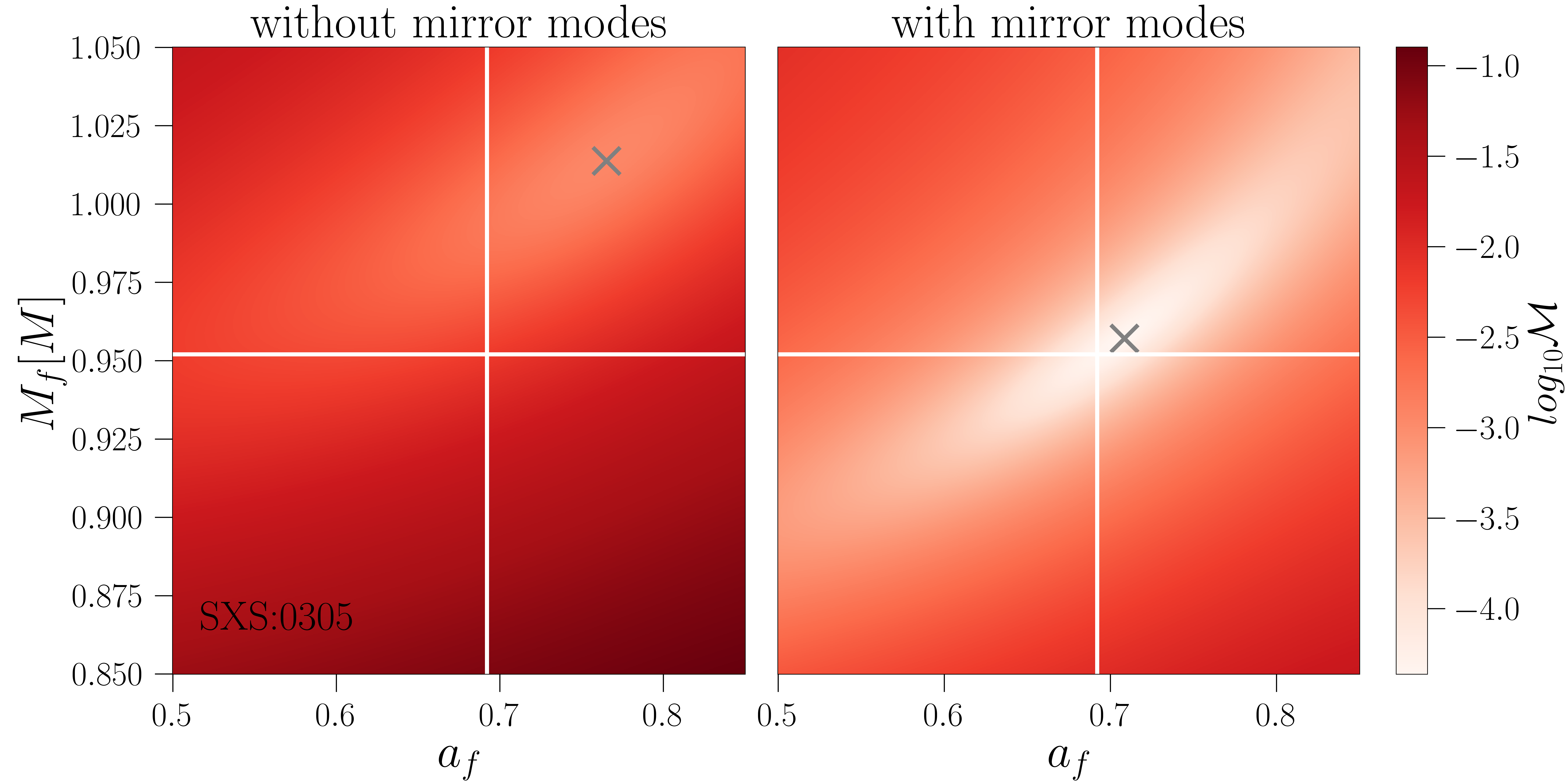}
    }
    \subfloat{
        \includegraphics[width=0.5\textwidth]{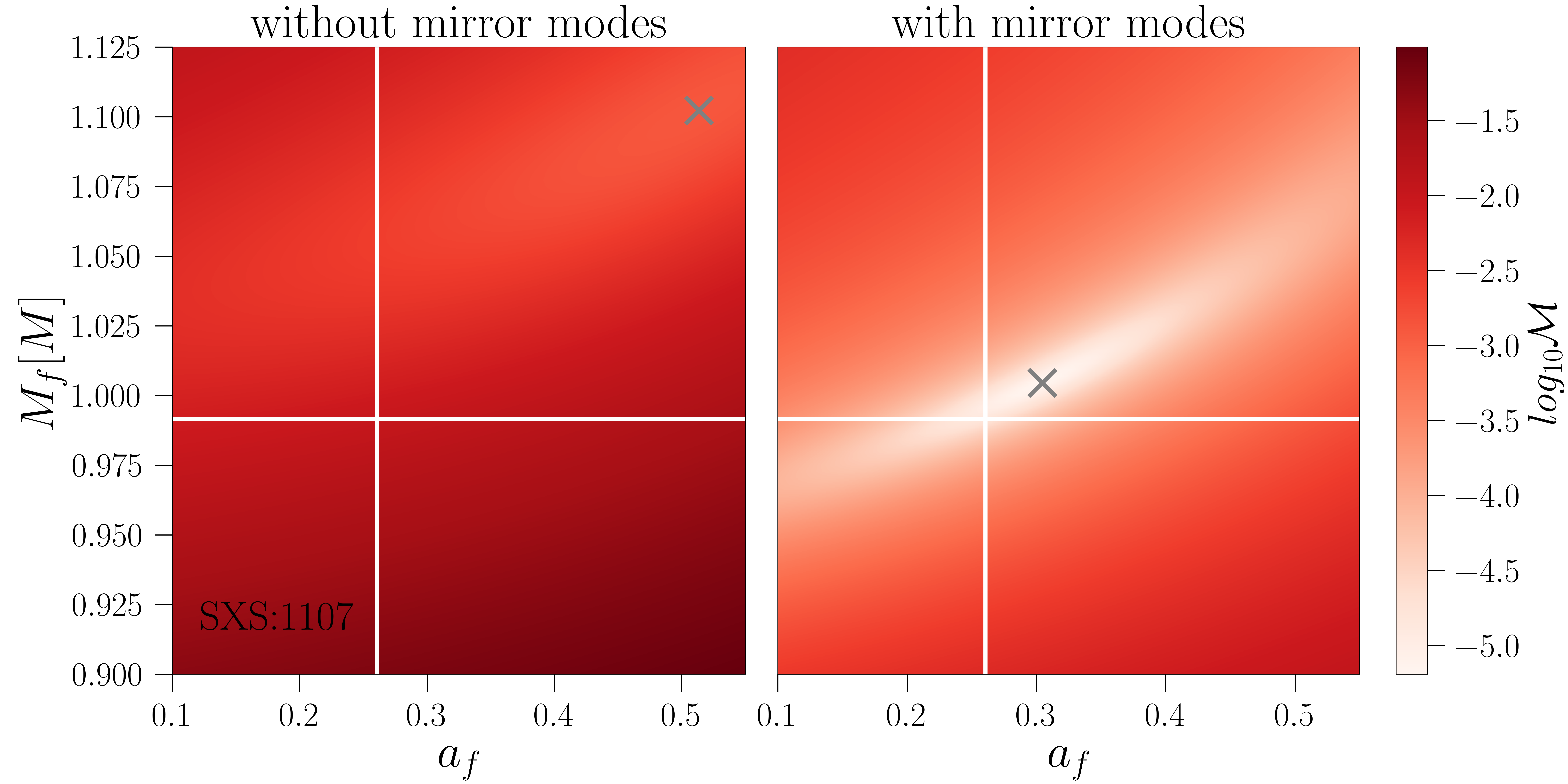}
    }
    \caption{Same as Fig.~\ref{fig:mismatch_grid_0} but the ringdown start time is taken to be $t_0=-10M$.
        It is clear that a linear 
        ringdown model including mirror modes recovers the true masses and spins much better
        at this time than the reference model.}
    \label{fig:mismatch_grid_-10}
\end{figure*}

As argued in \citet{PhysRevX.9.041060}, a ringdown model should not just
produce a small mismatch but also recover the correct physical parameters of the system. 
To this end, we vary $M_f$ and $a_f$ but keep the QNM frequencies to be that determined from 
perturbation theory (i.e., functions of $M_f$ and $a_f$) and repeat the mismatch calculation. 
A ringdown signal consisting of actual BH QNMs should minimize the mismatch 
for the true value of $M_f$ and $a_f$ as determined from NR simulations, modulo systematic errors.
A sharply peaked mismatch, on the other hand, would give better statistical errors
on the remnant paramters.
In Fig. \ref{fig:mismatch_grid_0} and Fig. \ref{fig:mismatch_grid_-10}, we plot the 
mismatch on a grid of $M_f$ and $a_f$ for two different start times, $t_0=0$ and $t_0=-10M$,
respectively, for an 8-tone model.
The left panel of each plot shows the heatmap of mismatches for the reference ringdown model and the 
right panel includes mirror modes in the ringdown model. 

We note that when the ringdown is started at the peak of $h_{22}$ mode,
the 8-tone reference model gives better estimates of the remnant parameters
than a model with mirror modes for $(l,m)=(2,2)$ while for $(l,m)=(2,1)$ mode, 
the remnant parameter estimates are roughly the same. 
But if we move the ringdown start time to an earlier fiducial time $t_0=-10M$, our
ringdown model has a deeper minimum near the correct remnant paramters. 
At this start time, the improvement in remnant estimates with the inclusion of mirror modes
is far greater for the large mass ratio binary than the (almost) equal mass one. 
This, as expected, points towards a greater significance of mirror modes for large mass ratios \citep{PhysRevD.73.064030}.
This aspect is also clear from the mismatch plots of the $(l,m)=(2,1)$ mode. 
The remnant parameter estimates using the $(l,m)=(2,1)$ mode is also far superior
with the inclusion of mirror mode ascertaining our earlier assertion that the mirror modes
excitation amplitudes depend on the value of $m$.

In Tab.~\ref{tab:mismatch}, we quote the value of the mismatch for the best-fit $M_f$ and $a_f$
when varied on a grid. Notice that the mirror mode model gives lower mismatches throughout with stark 
differences for the larger mass ratio case, $m=1$ mode, and starting time before $t_0=t_{\rm peak}$.
The only exception is the (almost) equal mass case where minimal excitation of mirror modes 
are expected and, therefore, by the time of the peak of the $h_{22}$ mode, these modes are no 
longer significant.

\begin{figure*}
    \centering
    \subfloat{
        \includegraphics[width=0.5\textwidth]{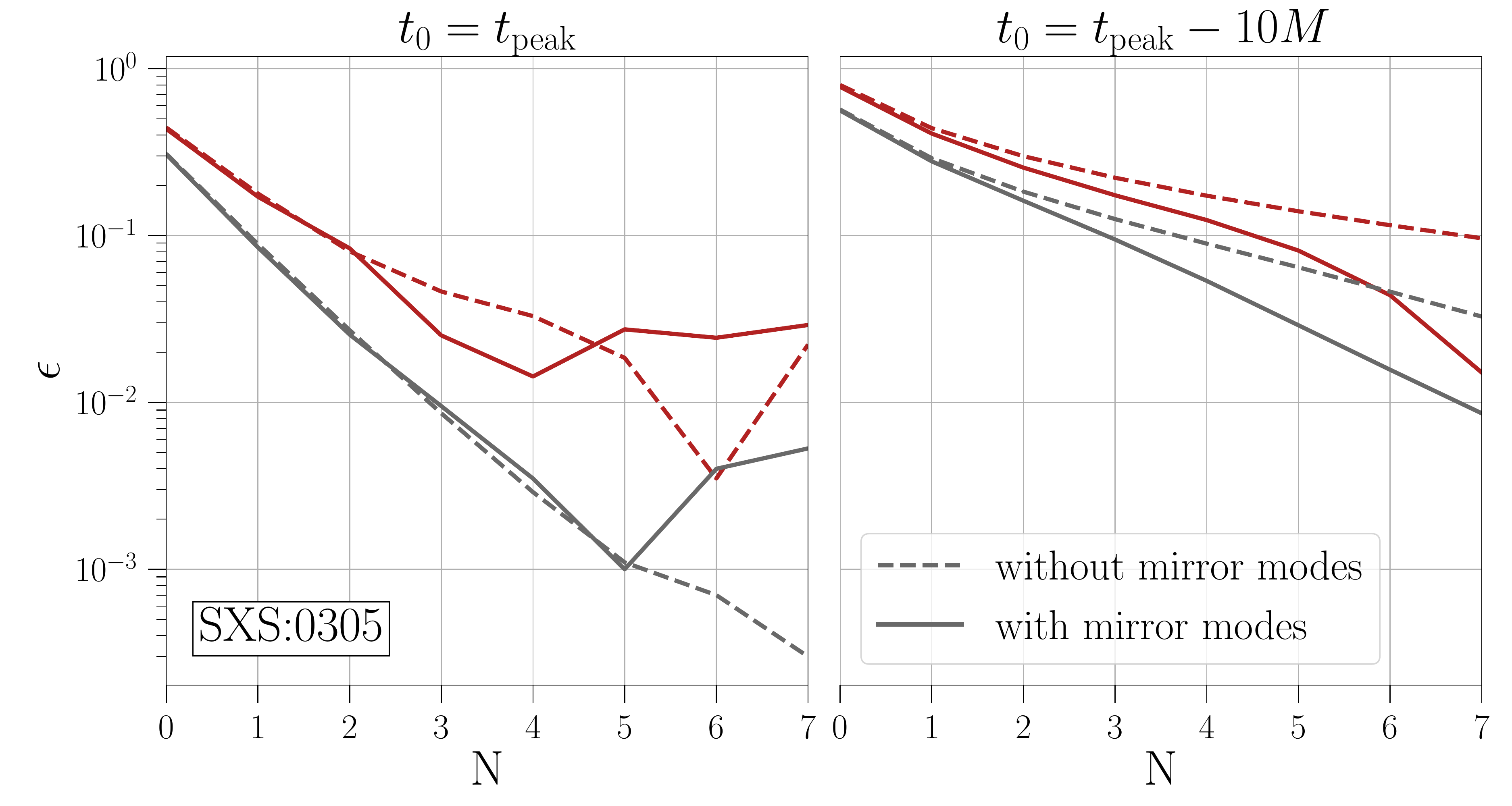}
    }
    \subfloat{
        \includegraphics[width=0.5\textwidth]{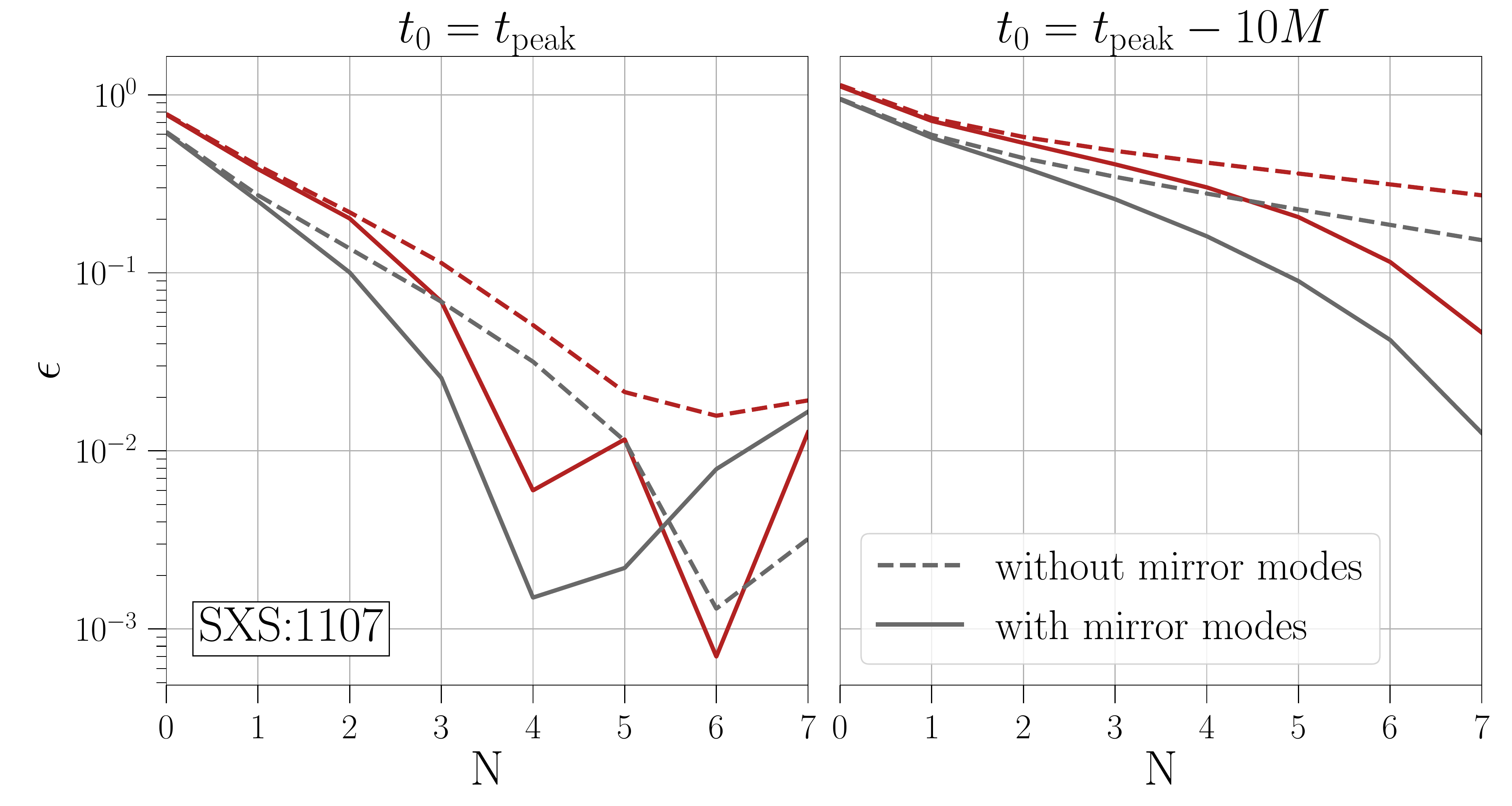}
    }
    \caption{$\epsilon$ as a function of the number of overtones included in a ringdown model.
        Solid (dashed) curves are for the mirror modes (reference) model. In the left two panels, 
        the ringdown is started at $t_0=t_{\rm peak}$ whereas on the right two panels, the ringdown
        start time is taken to be $t_0=t_{\rm peak}-10M$.}
    \label{fig:epsilon_0}
\end{figure*}

We quantify the errors in the estimation of the remnant parameters using a quantity $\epsilon$,
introduced in \citet{PhysRevX.9.041060}, defined as
\begin{equation}
    \epsilon = \sqrt{\left(\frac{\delta M_f}{M}\right)^2 + \left(\delta a_f\right)^2},
\end{equation}
where $\delta M_f$ and $\delta a_f$ are the differences between the best-fit values
and the true values of the remnant parameters as determined by NR. 
In Fig. \ref{fig:epsilon_0} we plot $\epsilon$ as a function of
the number of overtones in the ringdown model. 
We compare the performance of a ringdown model with mirror modes to that of the
reference model. We do the comparison at two different
start times, $t_0=t_{\rm peak}$ and $t_0=t_{\rm peak}-10M$. 
We see that when the ringdown is started at $t_0=t_{\rm peak}$, the mirror modes model performs as
good as or even marginally better than the reference model up to a 6-tone ringdown waveform for both the spherical harmonic modes
in the (almost) equal mass binary case. A higher-tone waveform model deteriorates the 
remnant parameter estimates for the mirror mode model.
For the high mass ratio case, the situation is different with the mirror mode model performing
much better than the reference model up to a 6-tone ringdown waveform for $m=2$ mode and 7-tone waveform for $m=1$ mode. 
If the ringdown in started at an earlier time $t_0=t_{\rm peak}-10M$, the mirror mode model
is clearly superior to the reference model for both the spherical harmonic modes and both 
the mass ratios under consideration. 
The trend is the same for both the modes and mass ratios, with the errors in the estimation of
remnant parameters decreasing monotonically with the number of included overtones and
the mirror mode model performing better by a factor of $\sim5-10$ for an 8-tone model. 

\begin{table}
    \centering
    \begin{ruledtabular}
        \begin{tabular}{c c c c c}
            & \multicolumn{2}{c}{SXS:0305} & \multicolumn{2}{c}{SXS:1107} \\

            \cline{2-3}
            \cline{4-5}

            $\textbf{m}$ & reference & mirror mode & reference & mirror mode \\

            \hline

            & \multicolumn{4}{c}{$t_0=t_{\rm peak}$} \\
            \cline{2-5}

            2 & $-6.51$ & $-6.84$ & $-6.15$ & $-6.62$ \\
            1 & $-3.96$ & $-5.34$ & $-3.79$ & $-6.71$ \\

            \hline

            & \multicolumn{4}{c}{$t_0=t_{\rm peak} - 10M$} \\
            \cline{2-5}

            2 & $-3.73$ & $-6.11$ & $-3.88$ & $-5.76$ \\
            1 & $-2.94$ & $-4.36$ & $-2.88$ & $-5.19$ \\

        \end{tabular}
    \end{ruledtabular}
    \caption{The value of the mismatch ($\log_{10}\mathcal{M}$) for the best-fit paramters in Fig.~\ref{fig:mismatch_grid_0}
        and Fig.~\ref{fig:mismatch_grid_-10}. Note that the mirror mode model gives orders of magnitude better mismatch for all the cases 
        except the (almost) equal mass case starting at $t_0=t_{\rm peak}$ where the mismatches are similar.}
    \label{tab:mismatch}
\end{table}

In Tab.~\ref{tab:amp22} and Tab.~\ref{tab:amp21}, we quote the real-valued amplitudes of regular modes ($\mathcal{A}_{lmn}$)
and mirror modes ($\mathcal{\bar{A}}_{lmn}$) for an 8-tone model for the two cases under study (SXS:0305 and SXS:1107) and
the two spherical harmonic modes $(l,m)=(2,2)$ and $(l,m)=(2,1)$, respectively.
The fit amplitudes are calculated at $t_0=-10M$.
We choose a fiducial reference time $t_0=0$ to quote the values of the amplitudes so that they can be easily
compared with the values quoted in other studies~\citep{PhysRevX.9.041060,Bhagwat:2019dtm}.
Eventhough the mirror mode amplitudes are much larger than the regular mode amplitudes at the start time $t_0=-10M$, 
we see that the regular modes become more dominant by the time the $h_{22}$ mode peaks.
At this time, the mirror modes are subdominant by close to an order of magnitude for most modes with up to $\sim 3$ orders of magnitude for $n=7$.
We point out that the amplitude of the positive frequency fundamental mode calculated at $t_0=-10M$ and time
evolved to $t_0=0$ is in striking agreement to that calculated at $t_0=0$ 
(and reported in \citet{PhysRevX.9.041060} and \citet{Bhagwat:2019dtm}). 
This indicates that the positive-frequency fundamental mode has entered the linear phase even at this early time. 
Additionally, we observe, for both the spherical harmonic modes, that the ratio between the mirror mode amplitudes 
and the corresponding regular mode amplitudes ($\mathcal{\bar{A}}_{lmn}/\mathcal{A}_{lmn}$)
is greater for the large mass ratio case indicating that mirror modes are
more strongly excited in large mass ratio binaries. 
Furthermore, note that, for both the mass ratios, this ratio is larger for $(l,m)=(2,1)$ mode than $(l,m)=(2,2)$ mode, 
demonstrating that mirror modes are more strongly excited in $m=1$ mode than $m=2$ mode.

We note the observation in~\citet{Bhagwat:2019dtm} that the best-fit amplitudes increase with the overtone
number $n$, reaches a maximum around $n=3/4$ and decreases for higher $n$ holds true for the mirror mode model as well
and as such provide support to their speculative reasoning that high-$n$ overtones are excited first by sources
far away from the horizon and hence are weaker. By contrast, low-$n$ overtones, excited by sources closer to the horizon,
falls partly into the horizon and does not reach null infinity. Consequently, intermediate-$n$ overtones are the most
strongly excited.

\begin{table}
    \centering
    \begin{ruledtabular}
        \begin{tabular}{c c c c c}
            & \multicolumn{4}{c}{$(l,m)=(2,2)$} \\
            \cline{2-5}
            & \multicolumn{2}{c}{SXS:0305} & \multicolumn{2}{c}{SXS:1107} \\

            \cline{2-3}
            \cline{4-5}

            $\textbf{n}$ & $\mathcal{A}_{lmn}$ & $\mathcal{\bar{A}}_{lmn}$ & $\mathcal{A}_{lmn}$ & $\mathcal{\bar{A}}_{lmn}$ \\

            \hline

            0 & 0.972455 & 0.00195254 & 0.337405 & 0.00104528 \\
            1 & 4.04150 & 0.0571118 & 1.10677 & 0.0409705 \\
            2 & 9.93874 & 0.535089 & 3.00238 & 0.503610 \\
            3 & 17.0806 & 1.43902 & 6.14453 & 2.15847 \\
            4 & 17.7877 & 0.672877 & 6.34204 & 2.32482 \\
            5 & 8.58776 & 0.0602107 & 2.59237 & 0.531559 \\
            6 & 1.43898 & 0.00153011 & 0.347826 & 0.0216096 \\
            7 & 0.0674631 & 1.37488$\mathrm{e}{-05}$ & 0.0106518 & 0.000144103 \\
        \end{tabular}
    \end{ruledtabular}
    \caption{Best-fit real amplitudes of $(l,m)=(2,2)$ mode for a mirror mode model with ringdown start time
        $t_0=t_{\rm peak}-10M$. The amplitude values quoted are at $t=t_{\rm peak}$ obtained by time-evolving 
        the respective amplitudes with their decay times. $\mathcal{A}_{lmn}$ are the amplitudes
        of the regular modes and $\mathcal{\bar{A}}_{lmn}$ are the mirror mode amplitudes.}
    \label{tab:amp22}
\end{table}

\begin{table}
    \centering
    \begin{ruledtabular}
        \begin{tabular}{c c c c c}
            & \multicolumn{4}{c}{$l=2$, $m=1$} \\
            \cline{2-5}
            & \multicolumn{2}{c}{SXS:0305} & \multicolumn{2}{c}{SXS:1107} \\

            \cline{2-3}
            \cline{4-5}

            $\textbf{n}$ & $\mathcal{A}_{lmn}$ & $\mathcal{\bar{A}}_{lmn}$ & $\mathcal{A}_{lmn}$ & $\mathcal{\bar{A}}_{lmn}$ \\

            \hline

            0 & 0.0528504 & 0.000369536 & 0.125076 & 0.00501953 \\
            1 & 0.326904 & 0.0217794 & 0.746708 & 0.0985377 \\
            2 & 1.08230 & 0.195302 & 2.67917 & 0.873092 \\
            3 & 2.13312 & 0.566340 & 5.84294 & 3.09317 \\
            4 & 2.13512 & 0.432410 & 5.73515 & 3.29656 \\
            5 & 0.921344 & 0.0753840 & 2.06885 & 0.913868 \\
            6 & 0.144722 & 0.00289362 & 0.236321 & 0.0535978 \\
            7 & 0.00606425 & 2.65232$\mathrm{e}{-05}$ & 0.00624921 & 0.000426323  \\
        \end{tabular}
    \end{ruledtabular}
    \caption{Same as Tab.~\ref{tab:amp22} but for $(l,m)=(2,1)$ mode.}
    \label{tab:amp21}
\end{table}

\begin{figure*}
    \centering
    \subfloat{
        \includegraphics[width=0.5\textwidth]{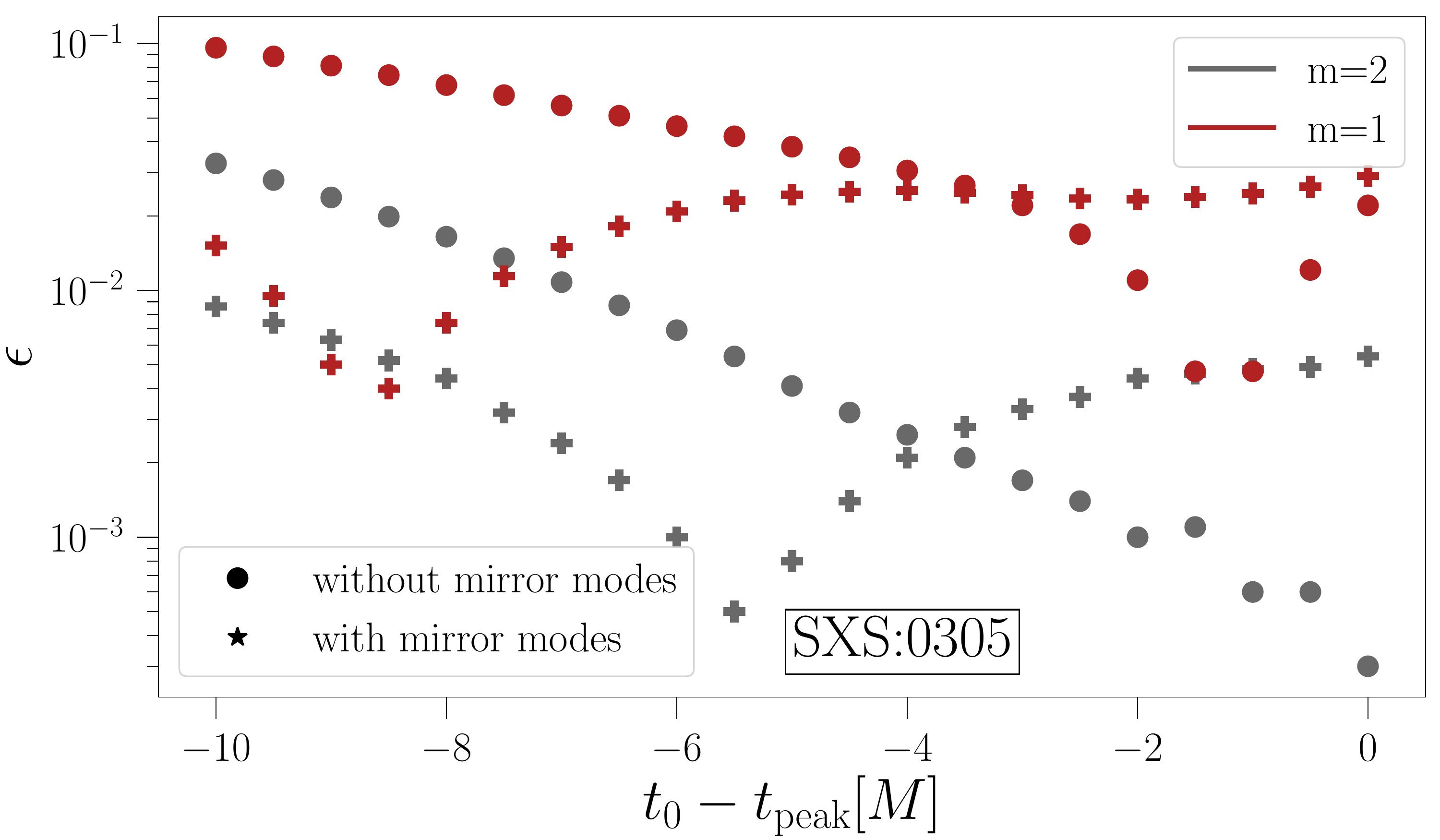}
    }
    \subfloat{
        \includegraphics[width=0.5\textwidth]{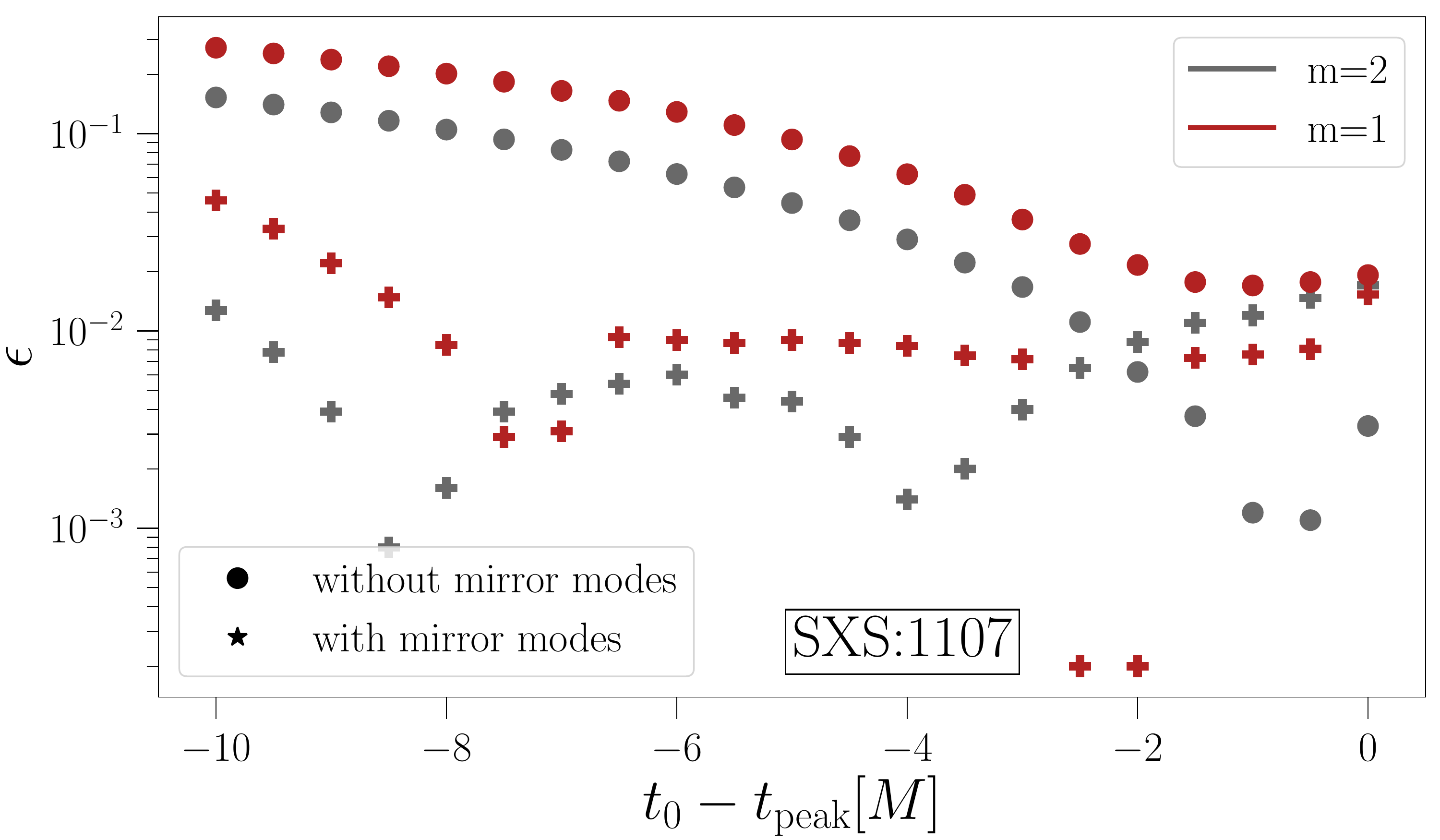}
    }
    \caption{$\epsilon$ as a function of the start time of ringdown $t_0$ for an 8-tone model. 
        Filled circles (pluses) show the value for reference (mirror mode) model. Note that 
        $\epsilon$ is consistently lower for the mirror mode model at early times for 
        both the spherical harmonic modes.}
    \label{fig:epsilon_t0}
\end{figure*}

Till this point, we have chosen a fiducial start time $t_0=t_{\rm peak}-10M$ to show the 
importance of mirror modes at times earlier than the peak of $h_{22}$ mode. We will now
look at the effect of a varying start time for an 8-tone ringdown model.
In Fig.~\ref{fig:epsilon_t0}, we show the error in the estimation of remnant parameters
as a function of the start time. 
We see that for the reference model, the errors increase (virtually) monotonically
the earlier the ringdown is started with respect to the peak of $h_{22}$.
For a mirror mode model, the error estimates have a minima at some time between
$t_0=t_{\rm peak}-10M$ and $t_0=t_{\rm peak}$ with the remnant parameter estimates 
using the mirror mode model being an order of magnitude more accurate. 
We observe that this minima occurs around the time that the mismatch curve
has a minima too. Therefore, we can also use the conventional wisdom of taking
the start time of ringdown at the earliest minima of a mismatch curve 
-- which is always at an earlier time for a mirror mode model than 
the corresponding minima for the reference model for a sufficiently high-tone ringdown waveform --
to argue the significance of mirror modes at times before the peak of $h_{22}$.
Furthermore, in the case of $(l,m)=(2,2)$ mode, the minimum errors in the 8-tone mirror mode model are about the same 
as that for the 8-tone reference model started at $t_0=t_{\rm peak}$ but 
with the advantage that the mirror mode model achieves this at a much earlier time, 
thereby accumulating more energy in the ringdown signal.
The situation is even better for $(l,m)=(2,1)$ mode where not only do the minimum
errors occur far before the peak of $h_{22}$ mode but also the errors are much smaller than 
that achieved by the reference model started at $t_0=t_{\rm peak}$.

We provide a speculative reasoning for why the remnant parameter estimates are better in the absence
of mirror modes when the ringdown is started near the peak of $h_{22}$ mode.
We reason this to be because mirror modes start well before the peak of $h_{22}$
-- we have seen that the fundamental positive frequency mode has already entered the 
ringdown phase at $-10M$ --
and due to their weaker excitation and shorter damping times compared to regular modes
(especially for the higher overtones where the difference in damping times becomes large), 
they decay to numerical noise by the time the $h_{22}$ mode peaks. 
This line of reasoning has support from Fig.~\ref{fig:epsilon_0} where one sees that
the inclusion of mirror modes improves parameter estimation for lower-tone models
mainly for the large mass ratio case where these negative-oscillation frequencies 
are excited more strongly. 
In principle, different modes should start at different times but allowing for this
in a pure ringdown model would introduce unphysical discontinuities in the waveform or 
waveform derivatives (see \citet{Bhagwat:2019dtm} for an expanded discussion). 
An option would be to attach higher order perturbation theory waveforms
to modes that start later so as to have a common earlier starting time for a higher order ringdown waveform
but that is beyond the scope of the current work.

\section{Conclusion and Discussion}
\label{sec:conclusion}

In this work, we studied the effects of including negative oscillation frequency modes in a ringdown waveform
which we call `mirror' modes. 
A ringdown signal from a non-spinning BBH merger has no apriori reason to consist of only corotating modes and, therefore,
mirror modes should be included in the ringdown waveform for a more accurate description of 
the gravitational wave signal.
We find that including mirror modes decreases the mismatch of our best-fit model with NR waveforms, with up to $\sim 3$ 
orders of magnitude improvement at times well before the peak of the $h_{22}$ mode. 
We further check whether the mismatches are minimized for the true values of mass and spin, if they are
allowed to vary, and find that the mirror mode model determines the remnant parameters better if 
the ringdown is started $6-9M$ before the peak of the $h_{22}$ mode. 
On the other hand, an 8-tone ringdown model
with only corotating modes gives better estimates of the remnant paramters if the 
ringdown is started at the peak of $h_{22}$ for the $(l,m)=(2,2)$ mode because mirror modes are not excited strongly
for the $m=2$ mode, although a 6-tone mirror mode model performs as good as a 6-tone reference model 
for the almost equal mass binary and an 8-tone reference model for the large mass ratio binary. 
We reason that the poorer performance of the mirror mode model when starting the ringdown at $t_0=t_{\rm peak}$
is because the mirror modes are excited at earlier times and they decay to numerical noise by the the time of the peak
of $h_{22}$ mode. 
We note that more work needs to be done in this regard to verify this claim. 
Having different start times for each mode would lead to unphysical discontinuities in the waveform or its derivatives
and, therefore, presents a technical challenge in pure ringdown modeling.
A possible route is to include second-order contributions and start the ringdown at an earlier time. This would ensure
smooth transition to linear regime for all the modes. 

A source of systematic that can affect our results is the use of mismatch as a quantifier for our fits.
It has been argued by \citet{Nollert:1999ji} and later by \citet{Berti:2007fi} that the fit-amplitudes of a mismatch-based 
approach cannot be regarded as the physical modes excited in the system. A better quantifier is 
Nollert's \emph{energy maximized orthogonal projection} (EMOP) that gives the energy parallel to a given QNM \citep{Nollert:1999ji,Berti:2007fi}.
We also point out that $\mathcal{M}$ as a function of the remnant parameters ($M_f$ and $a_f$) is an oscillatory function
with multiple local crests and troughs, which is an undesirable feature, not least because 
of the difficulty of locating the true remnant values.

If we trust the ringdown fits as the QNMs excited in the system, then it poses the question of what happened to all
the non-linearities present in the system? \citet{Bhagwat:2019dtm} argue that the conclusions of recent works to 
model the post-merger signal with a pure ringdown model is at odds with \citet{PhysRevD.97.104065}, where
the authors find appreciable non-linearities in the source frame near the common horizon. 
We bring to notice a more recent work of \citet{okounkova2020revisiting} that uses the same quantifiers of non-linearity as that
used in \citet{PhysRevD.97.104065}. The author then time evolves the gauge invariant quantifiers 
and finds that the non-linearities fall into the common horizon soon after its formation and does not 
reach asymptotic infinity. 

We believe further progress in ringdown modeling should take into account these findings.
Future work will include feasibility studies of detecting these mirror modes in LISA signals and in third generation
ground-based detectors.
We are also in the process of using this model to recover the remnant parameters for select events published by 
LIGO/Virgo.

\acknowledgements
I thank B. Sathyaprakash, Anuradha Gupta, Abhay Ashtekar, and E. Berti for useful discussions and 
B. Sathyaprakash and Anuradha Gupta for a careful reading of the manuscript.
I also thank Lionel London, Gregorio Carullo, and Vijay Varma for their comments on an initial draft of the 
manuscript.
I thank M. Giesler, M. Isi, and S. Bhagwat too for clarifications on their manuscripts \citet{PhysRevX.9.041060} 
and \citet{Bhagwat:2019dtm}.
I thank all front line workers combating the CoVID-19 pandemic without whose support this work would
not have been possible.

\bibliography{mirror_modes.bib}{}
\end{document}